\documentclass[letterpaper,12pt]{article}

\usepackage[left=1in,right=1in,top=1in,bottom=1in]{geometry}

\usepackage{setspace}
\usepackage[usenames]{color}
\usepackage[fleqn]{amsmath}
\usepackage{amssymb}
\usepackage{graphicx}
\usepackage{url}
\usepackage{verbatim}
\usepackage{appendix}
\usepackage{indentfirst}
\usepackage{booktabs}
\usepackage{multirow}
\usepackage[table, x11names]{xcolor}
\usepackage{ragged2e}
\usepackage{upgreek}
\usepackage{lscape}
\usepackage{longtable}
\usepackage[flushleft, referable]{threeparttablex}
\usepackage{rotating}
\usepackage[T1]{fontenc}
\usepackage[titles]{tocloft}
\usepackage{xspace}
\usepackage{ifthen}
\usepackage{cancel}
\usepackage{rotating}
\usepackage{array}
\usepackage{tabulary}
\usepackage{authblk}

\usepackage{hyperref}
\hypersetup{pdfborder={0 0 0}, colorlinks=true, urlcolor=black, linkcolor=black, citecolor=black}
\usepackage[capitalize]{cleveref}

\usepackage[right, mathlines]{lineno}
\usepackage{caption}
\DeclareCaptionLabelFormat{noSpace}{{#1}{#2}}
\DeclareCaptionListFormat{figList}{Figure {#2}.}
\DeclareCaptionListFormat{sFigList}{Figure S{#2}.}
\usepackage{subfig}

\usepackage[round]{natbib}
\usepackage{enumitem}
\input{macros.tex}

\newcommand{\keywords}{Approximate Bayesian computation; Bayesian model choice;
empirical Bayes; phylogeography; biogeography\xspace}

\title{Implications of uniformly distributed, empirically informed priors for
phylogeographical model selection: A reply to Hickerson et al.}

\author[1]{Jamie R.\ Oaks\thanks{Corresponding author: \href{mailto:joaks1@gmail.com}{\tt joaks1@gmail.com}}\thanks{Current address: Department of Biology, University of Washington, Seattle, Washington 98195}}
\author[2]{Charles W.\ Linkem}
\author[3]{Jeet Sukumaran}
\affil[1]{Department of Ecology and Evolutionary Biology, University of Kansas, Lawrence, Kansas 66045}
\affil[2]{Department of Biology, University of Washington, Seattle, Washington 98195}
\affil[3]{Department of Ecology and Evolutionary Biology, University of Michigan, Ann Arbor, Michigan 48109}

\date{\parbox{\linewidth}{\centering%
    \today\endgraf\bigskip
    \textbf{Running head}: Approximate Bayesian model choice\endgraf\bigskip
    \endgraf\bigskip
    \textbf{Counts}: 5403 words, 4 figures, 1 table, 10 supporting figures\endgraf\bigskip
    \textbf{Archival location}: \href{https://github.com/joaks1/msbayes-experiments}{\url{https://github.com/joaks1/msbayes-experiments}}}}

\makeatletter
\let\msTitle\@title
\let\msAuthor\@author
\let\msDate\@date
\makeatother

\begin{document}

\maketitle

\newpage

\begin{abstract}
    Establishing that a set of population-splitting events occurred at the same
time can be a potentially persuasive argument that a common process affected
the populations.
Recently, \citet{Oaks2012} assessed the ability of an approximate-Bayesian
model-choice method (\msb) to estimate such a pattern of simultaneous
divergence across taxa, to which \citet{Hickerson2013} responded.
Both papers agree that the primary inference enabled by the method is very
sensitive to prior assumptions and often erroneously supports shared
divergences across taxa when prior uncertainty about divergence times
is represented by a uniform distribution.
However, the papers differ about the best explanation and solution for this
problem.
\citet{Oaks2012} suggested the method's behavior was caused by the strong
weight of uniformly distributed priors on divergence times leading to smaller
marginal likelihoods (and thus smaller posterior probabilities) of models with
more divergence-time parameters (Hypothesis~\ref{hypML}); they proposed
alternative prior probability distributions to avoid such strongly weighted
posteriors.
\citet{Hickerson2013} suggested numerical approximation error causes \msb
analyses to be biased toward models of clustered divergences because the
method's rejection algorithm is unable to adequately sample the parameter space
of richer models within reasonable computational limits when using broad
uniform priors on divergence times (Hypothesis~\ref{hypError}).
As a potential solution, they proposed a model-averaging approach that uses
narrow, empirically informed uniform priors.
Here, we use analyses of simulated and empirical data to demonstrate that the
approach of \citet{Hickerson2013} does not mitigate the method's tendency to
erroneously support models of highly clustered divergences, and is dangerous in
the sense that the empirically-derived uniform priors often exclude from
consideration the true values of the divergence-time parameters.
Our results also show that the tendency of \msb analyses to support models of
shared divergences is primarily due to Hypothesis~\ref{hypML}, whereas
Hypothesis~\ref{hypError} is an untenable explanation for the bias.
Overall, this series of papers demonstrate that if our prior assumptions place
too much weight in unlikely regions of parameter space such that the exact
posterior supports the wrong model of evolutionary history, no amount of
computation can rescue our inference. 
Fortunately, as predicted by fundamental principles of Bayesian model choice,
more flexible distributions that accommodate prior uncertainty about parameters
without placing excessive weight in vast regions of parameter space with low
likelihood increase the method's robustness and power to detect temporal
variation in divergences.

\end{abstract}

\newpage

\section{Introduction}
Biogeographers frequently seek to explain population and species
differentiation on geographical phenomena.
Establishing that a set of population-splitting events occurred
at the same time can be a potentially persuasive argument that a set of taxa
were affected by the same geographic events.
The approximate-Bayesian method, \msb, allows biogeographers to estimate the
probabilities of models in which multiple sets of taxa diverge at the same
time \citep{Hickerson2006,Huang2011}.

Recently, \citet{Oaks2012} used this model-choice framework to study 22 pairs
of vertebrate lineages distributed across the Philippines; they also studied
the behavior of the \msb approach using computer simulations.
They found the method is very sensitive to prior assumptions and often
supports shared divergences across taxa that diverged randomly over broad time
periods, to which \citet{Hickerson2013} responded.
\citet{Oaks2012} and \citet{Hickerson2013} agree on the fundamental
methodological point about the model selection performed in \msb:
\begin{itemize}
   \item Representing prior uncertainty about divergence-time parameters with a
       uniform distribution can lead to spurious support for models with few
       divergence events shared across taxa. Thus, the primary inference
       enabled by the approach is very sensitive to the priors on divergence
       times.
\end{itemize}
However, the two papers suggest alternative mechanisms by which the priors on
divergence times cause this behavior:
\begin{enumerate}[label=\textsl{\textbf{Hypothesis \arabic*)}},ref=\arabic*,align=left]
    \item \textsl{\textbf{Strongly weighted marginal likelihoods}}
        \citep{Oaks2012}:
        The uniform priors on divergence times lead to very small marginal
        likelihoods (and thus smaller posterior probabilities) of models with
        many divergence-time parameters.  The likelihood of these models is
        ``averaged'' over a much greater parameter space in which there is a
        large amount of prior weight and small probability of producing the
        data.  \citep{Jeffreys1939,Lindley1957}. \label{hypML}
    \item \textsl{\textbf{Numerical-approximation error}} \citep{Hickerson2013}:
        Under broad uniform priors, the rejection algorithm implemented in \msb
        is unable to adequately sample the space of the models within
        reasonable computational time, which leads to bias toward models with
        fewer divergence-time parameters because they are better sampled.
        \label{hypError}
\end{enumerate}
In Hypothesis~\ref{hypError}, the problem is numerical-approximation error due
to insufficient computation.
In this scenario, given data from taxa that diverged randomly through time, the
exact (true) posterior supports a model with many divergence-time parameters,
but we are unable to accurately approximate this posterior.
In Hypothesis~\ref{hypML}, the problem is more fundamental; given data from
taxa that diverged randomly through time, the exact posterior supports a model
with simultaneous divergences across taxa.
I.e., when accommodating prior uncertainty about divergence times with a
uniform distribution, the exact posterior from Bayes' rule leads us to the
wrong conclusion about evolutionary history.
Such posterior support for simultaneous divergence, even if ``correct'' from
the perspective of Bayesian model choice, does not provide the biogeographical
insights that a researcher who employs \msb seeks to gain.

While these phenomena are not mutually exclusive, it is important to
distinguish between them in order to determine how to improve our ability to
estimate shared divergence histories.
If Hypothesis~\ref{hypError} is correct, then the model is sound and we need to
increase our computational effort or improve our Monte Carlo integration
procedures.
For example, Markov chain or sequential Monte Carlo algorithms might sample the
posterior more efficiently than the simple Monte Carlo rejection sampler
implemented in \msb.
Rather than alter the sampling algorithm, \citet{Hickerson2013} tried using
narrow, empirically informed uniform priors in the hope that with less
parameter space to sample, the rejection algorithm would produce better
estimates of the posterior.
Here, we discuss theoretical considerations for using empirically informed
priors for Bayesian model choice and evaluate the approach of
\citet{Hickerson2013} as a potential solution to the biases of \msb.
In their analyses, \citet{Hickerson2013} made an error by mixing different
units of time, which invalidates the results presented in their response (see
Supporting Information for details).
We correct this error, but still find their approach will often support
(1) clustered divergence models when divergences are random, and
(2) models that exclude from consideration the true values of the
parameters.

If Hypothesis~\ref{hypML} is correct, we need to correct the model, because no
amount of computation will help; even if we could calculate the exact
posterior, we would still reach the wrong interpretation about evolutionary
history.
Accordingly, \citet{Oaks2014dpp} has introduced a method that uses more
flexible probability distributions (e.g., gamma) to accommodate prior
uncertainty in divergence times without overly inhibiting the marginal
likelihoods of models with more divergence-time parameters.
This greatly increases the method's robustness and power to detect temporal
variation in divergences \citep{Oaks2014dpp}.
This is not surprising given the rich statistical literature showing that
marginal likelihoods are very sensitive to the priors used in Bayesian model
selection
\citep[e.g.,][]{Jeffreys1939,Lindley1957}.

We also use analyses of simulated and empirical data to explore the distinct
predictions made by Hypotheses 1 and 2.
We show the behavior of \msb matches the predictions of Hypothesis~\ref{hypML},
but not Hypothesis~\ref{hypError}.
This strongly suggests that the method tends to support models of shared
divergences not because of insufficient computation, but rather due to the
larger marginal likelihoods of these models under the prior assumption of
uniformly distributed divergence times.

\section{The potential implications of empirical Bayesian model choice}
\citet{Hickerson2013} suggest a very narrow, highly informed uniform prior on
divergence times is necessary to avoid the method's preference for models with
few divergence-time parameters.
Such an empirical Bayesian approach to model selection raises some theoretical
and practical concerns, some of which were discussed by \citet{Oaks2012} (see
the last paragraph of ``Assessing prior sensitivity of \msb'' in
\citet{Oaks2012}); we expand on this here.

\subsection{Theoretical implications of empirical priors for Bayesian model
choice}
\begin{linenomath}
Bayesian inference is a method of inductive learning in which Bayes' rule is
used to update our beliefs about a model $M$ as new information becomes
available.
If we let \allParameterValues represent the set of all possible parameter
values for model $M$, we can define a prior distribution for all $\theta \in
\allParameterValues$ such that $p(\theta \given M)$ describes our belief that
any given \myTheta{} is the true value of the parameter.
If we let \allDatasets represent all possible datasets then we can 
define a sampling model for all $\theta \in
\allParameterValues$ and $\alignment{}{} \in \allDatasets$ such that
$p(\alignment{}{} | \theta, M)$ measures our belief that any dataset \alignment{}{}
will be generated by any state \myTheta{} of model $M$.
After collecting a new dataset \alignment{i}{}, we can use Bayes' rule to
calculate the posterior distribution
\begin{equation}
    p(\myTheta{} \given \alignment{i}{}, M) = \frac{p(\alignment{i}{} \given
    \myTheta{}, M)p(\myTheta{} \given M)}{p(\alignment{i}{} \given M)},
    \label{eq:bayesrule}
\end{equation}
as a measure of our beliefs after seeing the new information, where
\begin{equation}
    p(\alignment{i}{} \given M) = \int_{\myTheta{}} p(\alignment{i}{} \given
    \myTheta{}, M)p(\myTheta{} \given M) d\myTheta{}
    \label{eq:marginallikelhiood}
\end{equation}
is the marginal likelihood of the model.
\end{linenomath}

This is an elegant method of updating our beliefs as data are accumulated.
However, this all hinges on the fact that the prior ($p(\myTheta{} \given M)$)
is defined for all possible parameter values independently of the new data
being analyzed.
Any other datasets or external information can safely be used to inform our
beliefs about $p(\myTheta{} \given M)$.
However, if the same data are used to both inform the prior and calculate the
posterior, the prior becomes conditional on the data, and Bayes' rule breaks
down.

Thus, empirical Bayes methods have an uncertain theoretical basis and
do not yield a valid posterior distribution from Bayes' rule \citep[e.g.,
empirical Bayesian estimates of the posterior are often too narrow, off-center,
and incorrectly shaped;][]{Morris1983,Laird1987,Carlin1990,Efron2013}.
This is not to say that empirical Bayesian approaches are not useful.
Empirical Bayes is a well-studied branch of Bayesian statistics that has given
rise to many methods for obtaining parameter estimates that often
exhibit favorable frequentist properties.
\citep{Morris1983,Laird1987,Laird1989, Carlin1990,Hwang2009}.

\begin{linenomath}
Although empirical Bayesian approaches can provide powerful methods for
parameter estimation, a theoretical justification for empirical Bayesian
approaches to model choice is questionable.
In Bayesian model choice, the primary goal is not to estimate parameters, but
to estimate the probabilities of candidate models.
In a simple example with two candidate models, $M_1$ and $M_2$, we can use
Bayes' rule to calculate the posterior probability of $M_1$ as
\begin{equation}
    p(M_1 \given \alignment{i}{}) = \frac{p(\alignment{i}{} \given
    M_1)p(M_1)}{p(\alignment{i}{} \given M_1)p(M_1) + p(\alignment{i}{} \given
    M_2)p(M_2)}.
    \label{eq:bayesmodelchoice}
\end{equation}
By comparing Equations \ref{eq:bayesrule} and \ref{eq:bayesmodelchoice}, we
see fundamental differences between Bayesian parameter estimation and
model choice.
\end{linenomath}

In Equation \ref{eq:bayesrule}, we see that the posterior density of any state
$\myTheta{}$ of the model, is the prior density updated by the probability of
the data given $\myTheta{}$ (the likelihood of $\myTheta{}$).
The marginal likelihood of the model only appears as a normalizing constant in
the denominator.
Thus, as long as the prior distribution contains the values of $\myTheta{}$
under which the data are probable and the data are strongly informative
relative to the prior, the values of the parameters that maximize the posterior
distribution will be relatively robust to prior choice, even if the posterior
is technically incorrect due to using the data to inform the priors.
However, if we look at Equation~\ref{eq:bayesmodelchoice}, we see that in
Bayesian model choice it is now the \emph{marginal} likelihood of a model that
updates the prior to yield the model's posterior probability.
The integral over the entire parameter space of the likelihood weighted by the
prior density is no longer a normalizing constant, rather it is how the data
inform the posterior probability of the model.
Because the prior probability distributions placed on the model's parameters
have a strong affect on the integrated, or ``average'', likelihood of a model,
Bayesian model choice tends to be much more sensitive to priors than parameter
estimation \citep{Jeffreys1939,Lindley1957}.
Another important difference of Bayesian model choice illustrated by
Equation~\ref{eq:bayesmodelchoice} is that the value of interest, the posterior
probability of a model, is not a function of \myTheta{} because the parameters
are integrated out of the marginal likelihoods of the candidate models.
Thus, unlike parameter estimates, the estimated posterior probability of a
model is a single value (rather than a distribution) lacking a measure of
posterior uncertainty.

The justification for an empirical Bayesian approach to parameter estimation is
that giving the data more weight relative to the prior (i.e., using the data
twice) will often shift the peak of the estimated distribution nearer to the
true value(s) of the model's parameter(s).
However, there is no such justification for model selection, because unlike
model parameters, the posterior probabilities of candidate models often have no
clear true values.
Model posterior probabilities are inherently measures of our belief in the
models after our prior beliefs are updated by the data being analyzed.
This complicates the meaning of model posterior probabilities when Bayes' rule
is violated by informing priors with the same data to be analyzed.
By using the data twice, we fail to account for prior uncertainty and mislead
our posterior beliefs in the models being compared; we will be over confident
in some models and under confident in others.

Nonetheless, empirical Bayesian model choice does perform well for some
problems.
Particularly, in cases where large aggregate data sets are used for many
parallel model-choice problems, pooling information to inform
priors can lead to favorable group-wise frequentist coverage across tests
\citep{Efron2008}.
However, this is far removed from the single model-choice problem of \msb.
In the Supporting Information we use a simple example to help highlight the
distinctions between Bayesian parameter estimation and model choice.

\subsection{Practical concerns about empirically informed uniform priors for
    Bayesian model choice}
In addition to the theoretical concerns discussed above, there are practical
problems with using narrow, empirically informed, uniform priors.
The results of Hickerson et al.'s (\citeyear{Hickerson2013}) reanalysis of the
Philippine dataset strongly favored models with the narrowest, empirically
informed prior on divergence times, and thus their model-averaged posterior
estimates are dominated by models $M_1$ and $M_2$ (see Table 1 of
\citet{Hickerson2013}).
This is concerning, because the narrowest \divt{} prior used by
\citet{Hickerson2013} ($\divt{} \sim U(0,0.1)$) likely excludes the true
divergence times for at least some of the Philippine taxa.
\citet{Hickerson2013} set this prior to match the 95\% highest posterior
density (HPD) interval for the mean divergence time estimated under one of the
priors used by \citet{Oaks2012} (see Tables 2 and 3 of \citet{Oaks2012}).
Given this interval estimate is for the \emph{mean} divergence time across all
22 taxa, it may be inappropriate to set this as the limit on the prior, because
some of the taxon pairs are expected to have diverged at times older than the
upper limit.
Furthermore, this prior is \emph{excluded} from the 95\% HPD interval estimates
of the mean divergence time under the other two priors explored by
\citet{Oaks2012} (under these priors the 95\% HPD is approximately 0.3--0.6;
see Table~6 of \citet{Oaks2012}).

The strong preference for the narrowest prior on divergence times suggests the
approach of \citet{Hickerson2013} is biased toward models with less parameter
space and, as a consequence, will estimate model-averaged posteriors dominated
by models that exclude true values of the parameters.
We explored this possibility in two ways.
First, we re-analyzed the Philippines dataset using the model-averaging
approach of \citet{Hickerson2013}, but set one of the prior models with a
uniform prior on divergence times that is unrealistically narrow and almost
certainly excludes most, if not all, of the true divergence times of the 22
taxon pairs.
If small likelihoods of large models cause the method to prefer models with
less parameter space (Hypothesis~\ref{hypML}), we expect \msb will
preferentially sample from this erroneous prior yielding a posterior that is
misleading (i.e., the model-averaged posterior will be dominated by a model
that excludes the truth).
Second, we generated simulated datasets for which the divergence times are
drawn from an exponential distribution and applied the approach of
\citet{Hickerson2013} to each of them to see how often the method excludes the
truth.

\subsubsection{Re-analyses of the Philippines dataset using empirical Bayesian
model averaging}

For our re-analyses of the Philippines dataset we followed the model-averaging
approach of \citet{Hickerson2013}, but with a reduced set of prior models to
avoid their error of mixing units of time (see SI for details).
We used five prior models, all of which had priors on population sizes of
$\meanDescendantTheta{} \sim U(0.0001, 0.1)$ and $\ancestralTheta{} \sim
U(0.0001, 0.05)$.
Following \citet{Hickerson2013}, each of these models had the following
priors on divergence times:
$M_1$, $\divt{} \sim U(0, 0.1)$;
$M_2$, $\divt{} \sim U(0, 1)$;
$M_3$, $\divt{} \sim U(0, 5)$;
$M_4$, $\divt{} \sim U(0, 10)$; and
$M_5$, $\divt{} \sim U(0, 20)$.
We simulated $1\e6$ random samples from each of the models for a total of
$5\e6$ prior samples.
For each model, we retained the 10,000 samples with the smallest Euclidean
distance from the observed summary statistics after standardizing the
statistics using the prior means and standard deviations of the given model.
From the remaining 50,000 samples, we then retained the 10,000 samples with the
smallest Euclidean distance from the observed summary statistics, this time
standardizing the statistics using the prior means and standard deviations
across all five models.
We then repeated this analysis twice, replacing the $M_1$ model with
$M_{1A}$ and $M_{1B}$, which differ only by having priors on divergence
times of $\divt{} \sim U(0, 0.01)$ and $\divt{} \sim U(0, 0.001)$,
respectively.
While we suspect the prior of $\divt{} \sim U(0, 0.1)$ used by
\citet{Hickerson2013} likely excludes the true divergence times of at least
some of the 22 taxa, we are nearly certain that these narrower priors exclude
most, if not all, of the divergence times of the Philippine taxa.

Our results show that the model-averaging approach of \citet{Hickerson2013}
strongly prefers the prior model with the narrowest distribution on divergence
times across all three of our analyses, even when this model excludes the true
divergence times of the Philippine taxa
(Table~\ref{tabModelChoiceEmpirical}).
Given that the same number of simuations were sampled from each prior model,
this behavior is not clearly predicted by insufficient computation
(Hypothesis~\ref{hypError}), but is a straightforward prediction of
Hypothesis~\ref{hypML}.

\citet{Hickerson2013} vetted the priors used in their model-averaging
approach via ``graphical checks,'' in which the summary statistics from 1000
random samples of each prior model are plotted along the first two orthogonal
axes of a principle component analysis (see Figure 1 of \citet{Hickerson2013}).
To determine if such prior-predictive analyses would indicate the $M_{1A}$ and
$M_{1B}$ models are problematic, we performed these graphical checks on our
prior models.
Unfortunately, these prior-predictive checks provide no warning that these
priors are too narrow (Figure~S\ref{figPCA}).
Rather, the graphs suggest these invalid priors are ``better fit''
(Figure~S\ref{figPCA}A--C) than the valid priors used by \citet{Oaks2012}
(Figure~S\ref{figPCA}D--F).

\subsubsection{Simulation-based assessment of Hickerson et al.'s
    (\citeyear{Hickerson2013}) model averaging over empirical priors}

To better quantify the propensity of Hickerson et al.'s
(\citeyear{Hickerson2013}) approach to exclude the truth, we simulated 1000
datasets in which the divergence times for the 22-population pairs are drawn
randomly from an exponential distribution with a mean of 0.5 ($\divt{} \sim
Exp(2)$).
All other parameters were identically distributed as the $M_1$--$M_5$ models
(Table~\ref{tabModelChoiceEmpirical}).
We then repeated the model-averaging  analysis described above, retaining 1000
posterior samples for each of the 1000 simulated datasets.
For each simulation replicate, we estimated the Bayes factor in favor
of excluding the truth as the ratio of the posterior to prior odds of
excluding the true value of at least one parameter.
Whenever the Bayes factor preferred a model excluding the truth, we counted the
number of the 22 true divergence times that were excluded by the preferred
model.

Our results show that the model-averaging approach of \citet{Hickerson2013}
favors a model that excludes the true values of parameters in 97\% of the
replicates (90\% with GLM-regression adjustment), excluding up to 21 of the 22
true divergence times (Figure~\ref{figExclusionSimTau}).
Importantly, the posterior probability of excluding at least one true parameter
value is very high in most replicates
(Figure~\ref{figExclusionSimProb}).
Using a Bayes factor of greater than 10 as a criterion for strong support, 66\%
of the replicates (87\% with GLM-regression adjustment) strongly support the
exclusion of true values (Figure~\ref{figExclusionSimProb}).

The results of the above empirical and simulation analyses clearly demonstrate
the risk of using narrow, empirically guided uniform priors in a Bayesian
model-averaging framework.
The consequence of this approach is obtaining a model-averaged posterior
estimate that is heavily weighted toward models that exclude true
values of the parameters.
This is not a general critique of Bayesian model averaging.
Rather, model averaging can provide an elegant way of incorporating
model uncertainty in Bayesian inference.
However, as predicted by Hypothesis~\ref{hypML}, when averaging over models
with narrow and broad uniform priors on a parameter that is not expected to
have a uniformly distributed likelihood density, the posterior can be dominated
by models that exclude from consideration the true values of parameters due to
their larger marginal likelihoods (these models integrate over less space with
high prior weight and low likelihood).

When using uniformly distributed priors, the alternative to capturing prior
uncertainty is to risk excluding the true values one seeks to estimate.
Fortunately, more flexible continuous distributions that are better suited as
priors for the positive real-valued parameters of the \msb model have been
shown to greatly reduce spurious support for clustered divergence models while
allowing prior uncertainty to be accommodated
\citep{Oaks2014dpp}.

\section{Assessing the power of the model-averaging approach of
    \citet{Hickerson2013}}
While our results above clearly demonstrate the risks inherent to the empirical
Bayesian model-choice approach used by \citet{Hickerson2013}, one could justify
such risk if the approach does indeed increase power to detect temporal
variation in divergences.
We assess this possibility using simulations.
Following \citet{Oaks2012}, we simulated 1000 datasets with \divt{} for each of
the 22 population pairs randomly drawn from a uniform distribution, $U(0,
\divt{max})$, where \divt{max} was set to: 0.2, 0.4, 0.6, 0.8, 1.0, and 2.0, in
\globalcoalunit generations.
All other parameters were identically distributed as the prior models.
As above, we generated $5\e6$ samples from prior models $M_1$--$M_5$
(Table~\ref{tabModelChoiceEmpirical}).
For each of the 6000 simulated datasets, we approximated the posterior
by retaining 1000 samples from the prior.

Our results demonstrate that the approach of \citet{Hickerson2013} consistently
infers highly clustered divergences across all the \divt{max} we simulated
(Figure~\ref{figPower4}A--D \& S\ref{figPower6}A--F).
The approach often strongly supports (Bayes factor of greater than 10) the
extreme case of one divergence event across all our simulation conditions
(Figure~\ref{figPower4}E--H \& S\ref{figPower6}G--L).
The method also struggles to estimate the variance of divergence times
(\vmratio{}), whether evaluating the unadjusted
(Figure~S\ref{figPowerAccuracy}A--F) or GLM-adjusted
(Figure~S\ref{figPowerAccuracy}G--L) posterior estimates.
Overall, the empirical Bayesian model-averaging approach leads to erroneous
support for highly clustered divergences when populations diverged randomly
over the last $8\globalpopsize$ generations.
For loci with per-site rates of mutation on the order of $1\e{-8}$ and
$1\e{-9}$ per generation, this translates to 10 million and 100 million
generations, respectively.

Also, the results of our power analyses further demonstrate the propensity of
Hickerson et al.'s (\citeyear{Hickerson2013}) approach to exclude true
parameter values.
Across all but one of the \divt{max} we simulated, the method favors a model
that excludes the truth in a large proportion of replicates, and across many of
the \divt{max} the preferred model will exclude a large proportion of the true
divergence times (Figure~\ref{figPowerExclusion4}A--D \&
S\ref{figPowerExclusion6}A--F).
Importantly, the posterior probability of excluding at least one true
divergence value is also quite high across many of the \divt{max}
(Figure~\ref{figPowerExclusion4}E--H \& S\ref{figPowerExclusion6}G--L).

\section{The importance of power analyses to guide applications of \texttt{msBayes}}
\citet{Hickerson2013} presented a power analysis of \msb under a narrow uniform
divergence-time prior of 0--1 coalescent units ago.
They found that under these prior conditions \msb can, assuming a
per-site rate of $1.92\e{-8}$ mutations per generation, detect multiple
divergence events among 18 taxa when the true divergences were random over
150,000 generations or more.
It is important that investigators perform such simulations to determine the
method's power for their dataset, and decide if \msb has sufficient temporal
resolution to address their hypotheses; in the case of the Philippines dataset,
it did not.
When doing so, it is important to consider what prior conditions are relevant
to the empirical system.
It is rare for there to be enough \emph{a priori} information to be certain
that all taxa diverged within the last 4$\globalpopsize$ generations (i.e.,
0--1 coalescent units).
Also, it seems unlikely that when such prior information is available that
being able to detect more than one divergence event in the face of 18
divergences that were random over 150,000+ generations will
provide much insight into the evolutionary history of the taxa.

Inferring more than one divergence time shared across all taxa does not confirm
the method is working well when analyzing data generated under random temporal
variation in divergences (e.g., an inference of two divergence events could be
biogeographically interesting yet spurious).
Thus, it is important that investigators not limit their assessment of the
method's power to only differentiating inferences of one event or more (i.e.
$\numt{} = 1$ versus $\numt{} > 1$).
Rather, looking at the distribution of estimates, as in Figure~\ref{figPower4}
and \citet{Oaks2012}, provides much more information about the behavior of the
method.

\section{The causes of support for models of co-divergence}
To determine how best to improve the behavior of \msb, it is important to
determine the mechanism by which broad uniform priors cause support for
clustered models of divergence.
It is well established that vague priors can be problematic in Bayesian model
selection.
Models that integrate over more parameter space characterized by low
probability of producing the data and relatively high prior density will have
smaller marginal likelihoods \citep{Jeffreys1939,Lindley1957}.
Given the uniformly distributed priors on divergence times employed in \msb,
the likelihood of models with more divergence parameters will be ``averaged''
over much greater parameter space, all with equal prior weight, and much of it
with small likelihood (Hypothesis~\ref{hypML}).
In light of this fundamental statistical issue, it is not surprising that the
method tends to support simple models.

However, \citet{Hickerson2013} conclude that the bias is caused by numerical-
approximation error due to insufficient computation
(Hypothesis~\ref{hypError}).
They argue the widest of the three priors on divergence times used by
\citet{Oaks2012} would infrequently produce random samples of parameter values
with many independent population divergence times as recent as the estimated
gene divergence times presented in \citet{Oaks2012}.
However, this sampling-probability argument is based on some questionable
assumptions.
Oaks et al.'s (\citeyear{Oaks2012}) gene-tree estimates were intended to
provide only a rough comparison of the gene divergence times across the 22 taxa
and assumed an arbitrary strict per-site rate of $2\e{-8}$ mutations per
generation for all taxa.
Furthermore, because the branch-length units of the gene trees are in millions
of years whereas the divergence-time prior of \msb is in generations,
\citet{Hickerson2013} make the implicit assumption that all 22 Philippine taxa
have a generation time of one year.
More importantly, even if we assume (1) the arbitrary strict clock is correct,
(2) gene divergence times were estimated without error, and (3) all 22 taxa
have one-year generation times, Hickerson et al.'s (\citeyear{Hickerson2013})
argument actually demonstrates that the models used by \citet{Oaks2012} with
narrower priors on divergence times are densely populated with samples with
large numbers of divergence parameters with values younger than the estimated
gene divergence estimates.
Thus, if \citet{Hickerson2013} are correct, analyses under these narrow priors
should be much less biased toward clustered models of divergence.
However, the magnitude of the bias is very similar across all three priors
explored by \citet{Oaks2012}.
\citet{Hickerson2013} point out a case where the narrowest prior performs
slightly better (panel L of Figures S32, S37, and S38 of \citet{Oaks2012}).
However, it is important to note that these results suffered from a bug in
\msb, and after \citet{Oaks2012} corrected the bug, there are many cases where
the narrowest prior performs slightly worse (see panels D--J of Figures 3 and
S12 of \citet{Oaks2012}).

To disentangle whether Hypothesis 1 or 2 is the primary cause of the method's
erroneous support for simple models, we must look at the different predictions
made by these two phenomena.
For example, numerical error due to insufficient prior sampling
(Hypothesis~\ref{hypError}) should create large variance among posterior
estimates and cause analyses to be highly sensitive to the number of samples
drawn from the prior.
Furthermore, if insufficient prior sampling is \emph{biasing} estimates toward
models with less parameter space we expect to see support for these models
decrease as sampling from the prior increases.
\citet{Oaks2012} did not see such sensitivity when they compared prior sample
sizes of $2\e6$, $5\e6$, and $10^7$.

To explore this prediction further, we repeat the analysis of the Philippines
dataset under the intermediate prior used by \citet{Oaks2012} ($\divt{} \sim
U(0, 10)$, $\meanDescendantTheta{} \sim (0.0005, 0.04)$, $\ancestralTheta{}
\sim (0.0005, 0.02)$), using a very large prior sample size of $10^8$.
When we look at the trace of the estimates of the dispersion index of
divergence times (\vmratio{}) as the prior samples accumulate
(Figure~S\ref{figSamplingError}) we do not see the trend predicted by
Hypothesis~\ref{hypError}.
While approximation error is always present in any numerical analysis, it does
not appear to be playing a large role in the biases revealed by the results of
\citet{Oaks2012} or presented above.

A straightforward prediction if strongly weighted marginal likelihoods are
causing the preference for simple models (Hypothesis~\ref{hypML}) is that the
bias should disappear as the model generating the data converges to the prior.
\citet{Oaks2012} tested this prediction by performing 100,000 simulations to
assess the model-choice behavior of \msb when the prior model is correct.
The results confirm the prediction of Hypothesis~\ref{hypML}:
\msb estimates the probability of the one-divergence model quite well (or even
\emph{under}estimates it) when the prior is correct (see Figure 4 of
\citet{Oaks2012}).
We confirmed this same behavior for the model-averaging approach used by
\citet{Hickerson2013} (see SI text and Figure~S\ref{figValidationMCBehavior}).
These results are not clearly predicted if insufficient computation was causing
numerical error (Hypothesis~\ref{hypError}).
Even when the prior is correct, due to the discrete uniform prior on the number
of divergence events (\numt{}) implemented in \msb, models with larger numbers
of divergence-time parameters (and thus greater parameter space) will still be
far less densely sampled than those with fewer divergence events
\citep{Oaks2012}.
Thus, the results of the simulations of \citet{Oaks2012} are more consistent
with the fundamental sensitivity of marginal likelihoods to priors
(Hypothesis~\ref{hypML}).

This is further demonstrated by the results presented herein that show the
model-averaging approach of \citet{Hickerson2013} prefers models with narrower
\divt{} priors (Table~\ref{tabModelChoiceEmpirical} and
Figs.~\labelcref{figExclusionSimTau,figExclusionSimProb,figPowerExclusion4})
and fewer \divt{} parameters (Figure~\ref{figPower4}).
For these model-averaging analyses, insufficient prior sampling
(Hypothesis~\ref{hypError}) is an untenable explanation for the erroneous support
for models with less parameter space, because (1) all of the prior models share
the same dimensionality, and (2) the same number of random samples were drawn
from each of the prior models.
However, these results are predicted by Hypothesis~\ref{hypML}, because the
marginal likelihoods will be higher for models with narrower priors on
divergence times and fewer divergence-time dimensions (these models integrate
over less space with large prior weight and small likelihood).

\section{Improving inference of shared divergences}
In theory, the model-averaging approach of \citet{Hickerson2013} is appealing.
It leverages a great strength of Bayesian statistical procedures, namely the
ability to obtain marginalized estimates that incorporate uncertainty in
nuisance parameters.
However, when sampling over models with narrow-empirical and diffuse uniform
priors for a parameter that is expected to have a very non-uniform likelihood
density, models that exclude the true values of the parameters we aim to
estimate will often have the largest marginal likelihoods.

The recommendations of \citet{Oaks2012} for mitigating the lack of robustness
of \msb are similar to those of \citet{Hickerson2013}, but avoid the need for
imposing an additional dimension of model choice and using priors that often
exclude the truth.
\citet{Oaks2012} suggest that uniform priors may not be ideal for many
parameters of the \msb model, and recommend the use of probability
distributions from the exponential family.
If we look at the prior distribution on divergence times imposed by
the model-averaging approach of \citet{Hickerson2013} we see it is a mixture of
overlapping uniforms with lower limits of zero (Figure~S\ref{figMCTauPrior}).
This looks very much like an exponential distribution, except that in any state
of the model, all the divergence times are restricted to the hard
bounds of one of the uniform distributions.
Thus, it seems more appropriate to simply place a gamma prior (the exponential
being a special case) on divergence times.
This would capture the prior uncertainty that \citet{Hickerson2013} are
suggesting for divergence times (Figure~S\ref{figMCTauPrior}) while avoiding
costly model-averaging and the constraint that all divergence times must fall
within the hard bounds of the current model state.
It also would allow an investigator to place the majority of the prior density
in regions of parameter space they believe, \emph{a priori}, are most
plausible, but still capture uncertainty in the tails of distributions with low
density.
Indeed, \citet{Oaks2014dpp} has shown that the use of gamma distributions in
place of uniform priors improves the power of the method to detect temporal
variation in divergences and reduces erroneous support for clustered
divergences.

\section{Conclusions}
We demonstrate how the approximate-Bayesian model-choice method implemented in
\msb can spuriously support models with less parameter space.
This is caused by the use of uniform priors on divergence times.
Uniform distributions necessitate the use of priors that place high density in
unlikely regions of parameter space, less the risk of excluding the true
divergence times \emph{a priori}.
These broad uniform priors reduce the marginal likelihoods of models with more
divergence-time parameters.
We show that the empirical Bayesian model-averaging approach of
\citet{Hickerson2013} does not mitigate this bias, but rather causes it to
manifest by sampling predominantly from models that often exclude the true
values of the divergence times.
Our results show that it is difficult to choose an uniformly distributed prior
on divergence times that is broad enough to confidently contain the true values
of parameters while being narrow enough to avoid strongly weighted and
misleading posterior support for models with less parameter space.
More generally, it is important to carefully choose prior assumptions about
parameters in Bayesian model selection, because they can strongly influence the
posterior probabilities of the models we seek to compare.
No amount of computation can rescue our inference if our prior assumptions
place too much weight in unlikely regions of parameter space such that the
exact posterior supports the wrong model of evolutionary history. 

The common inference of temporally clustered historical events
\citep{Barber2010, Bell2012, Carnaval2009, Chan2011, Chan2014, Daza2010,
    Hickerson2006, Huang2011, Lawson2010, Leache2007, Plouviez2009, Stone2012,
    Voje2009},
when not accompanied with the necessary analyses to assess the robustness and
temporal resolution of such results, should be treated with caution, because
\msb has been shown to erroneously infer clustered events over a range of prior
conditions.
Fortunately, \citet{Oaks2014dpp} has shown that alternative probability
distributions allow prior uncertainty to be accommodated while avoiding
excessive prior density in regions of low likelihood, which greatly improves
inference of shared divergence histories.

The work presented herein follows the principles of Open Notebook Science.
All aspects of the work were recorded in real-time via version-control software
and are publicly available at
\href{https://github.com/joaks1/msbayes-experiments}{\url{https://github.com/joaks1/msbayes-experiments}}.
All information necessary to reproduce our results is provided there.

\section{Acknowledgments}
We thank Melissa Callahan, Jake Esselstyn, Cameron Siler, Mark Holder, Rafe
Brown, Emily McTavish, Daniel Money, Jordan Koch, Adam Leach\'{e}, Vladimir
Minin, Luke Harmon, and three anonymous reviewers for insightful comments that
greatly improved this work.
We thank Michael Hickerson and co-authors for generously providing their data.
J.\ Oaks and C.\ Linkem thank the National Science Foundation for supporting
this work (DEB 1011423, DBI 1308885 and BIO-1202754).
J.\ Oaks was also supported by the University of Kansas (KU) Office of Graduate
Studies, Society of Systematic Biologists, Sigma Xi Scientific Research
Society, KU Department of Ecology and Evolutionary Biology, and the KU
Biodiversity Institute.
We also thank Mark Holder, the KU Information and Telecommunication Technology
Center, KU Computing Center, and the iPlant Collaborative for the computational
support necessary to conduct the analyses presented herein.

\bibliographystyle{evolution.bst}
\bibliography{references}

\newpage

\renewcommand\listfigurename{Figure Captions}
\cftsetindents{fig}{0cm}{2.2cm}
\renewcommand\cftdotsep{\cftnodots}
\setlength\cftbeforefigskip{10pt}
\cftpagenumbersoff{fig}
\listoffigures


\newpage
\singlespacing

\begin{table}[htbp]
    \sffamily
    \addtolength{\tabcolsep}{-0.08cm}
    \rowcolors{2}{}{myGray}
    \caption{Results of the model-averaging approach of \citet{Hickerson2013}
        applied to the Philippines dataset of \citet{Oaks2012} using three sets
        of prior models. All models used priors on population size of
        $\meanDescendantTheta{} \sim U(0.0001,0.1)$ and $\ancestralTheta{} \sim
        U(0.0001, 0.05)$, and differ only in their prior on divergence-time
        (\divt{}) parameters.  Each set of five models differ only in the
        divergence-time prior used for the model with the narrowest prior:
        $M_1$ ($\divt{} \sim U(0, 0.1)$), $M_{1A}$ ($\divt{} \sim U(0, 0.01)$),
        or $M_{1B}$ ($\divt{} \sim U(0, 0.001)$). The approximate posterior
        probability of each model ($p(M_i \given \ssSpace)$) is given for each
        of the three analyses.  The posterior estimates are based on 10,000
        samples retained from $1\e6$ prior samples
    from each model.}
    \centering
    \begin{tabular}{ l l l l l }
        \toprule
        & & \multicolumn{3}{c}{$p(M_i \given \ssSpace)$} \\
        \cmidrule(){3-5}
        Model & \divt{} prior & $M_{*}=M_1$ & $M_{*}=M_{1A}$ & $M_{*}=M_{1B}$ \\
        \midrule
        $M_*$ & --        & 0.899 & 0.821 & 0.673 \\
        $M_2$ & $U(0,1)$  & 0.079 & 0.136 & 0.251 \\
        $M_3$ & $U(0,5)$  & 0.013 & 0.026 & 0.044 \\
        $M_4$ & $U(0,10)$ & 0.006 & 0.012 & 0.022 \\
        $M_5$ & $U(0,20)$ & 0.003 & 0.005 & 0.010 \\
        \bottomrule
    \end{tabular}
    \label{tabModelChoiceEmpirical}
\end{table}

\clearpage

\newpage

\mFigure{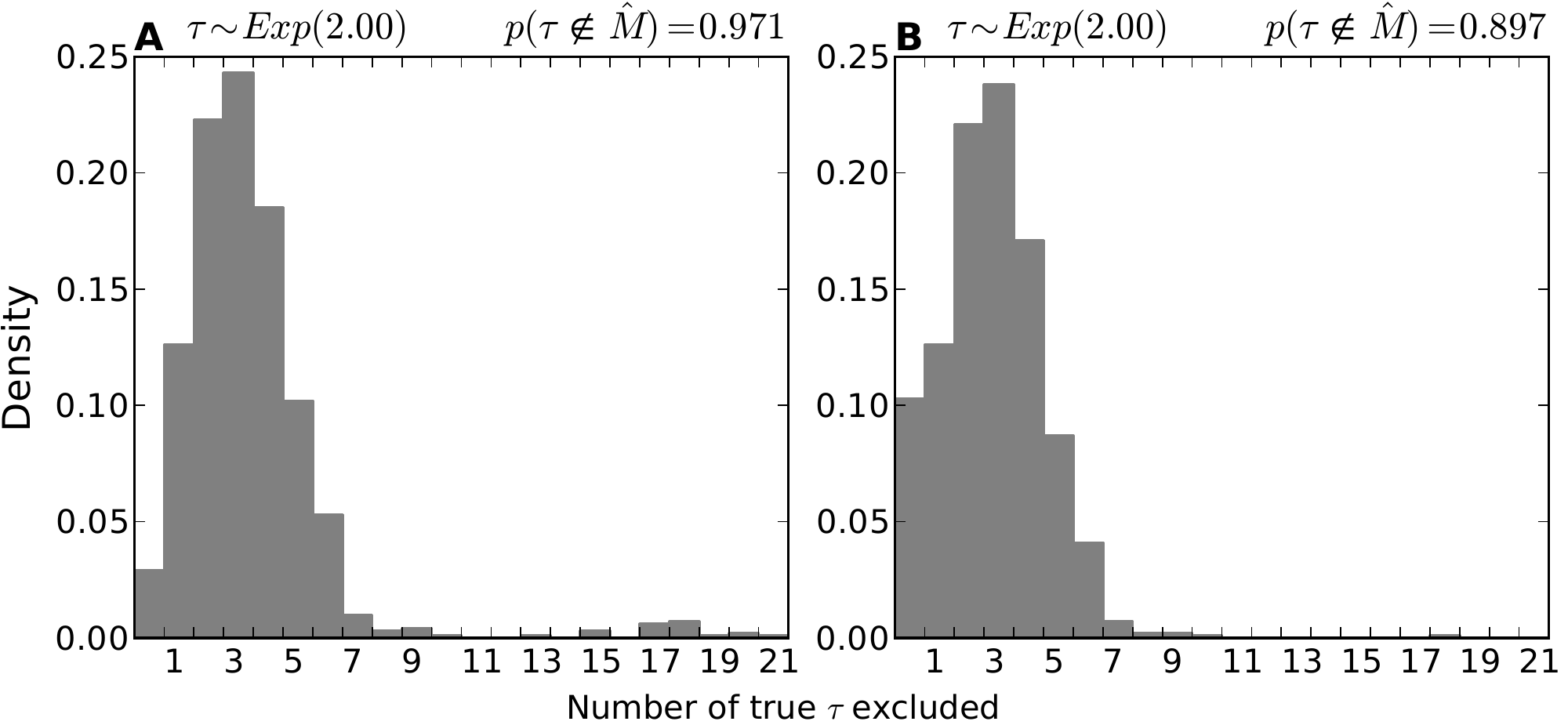}{
    Histograms of the number of true divergence times excluded from the model
    preferred by the empirically informed model-averaging approach of
    \citet{Hickerson2013} when applied to simulated datasets in which divergence
    times of 22 pairs of populations are drawn from an exponential
    distribution, $\divt{} \sim Exp(2)$.
    The plots represent (A) unadjusted and (B) GLM-adjusted estimates from 1000
    simulation replicates analyzed using $5\e6$ samples from the prior.
    The proportion of simulation replicates in which at least one true
    parameter value is excluded from the preferred model ($p(\divt{} \notin
    \hat{M})$) is also given.
}{figExclusionSimTau}

\mFigure{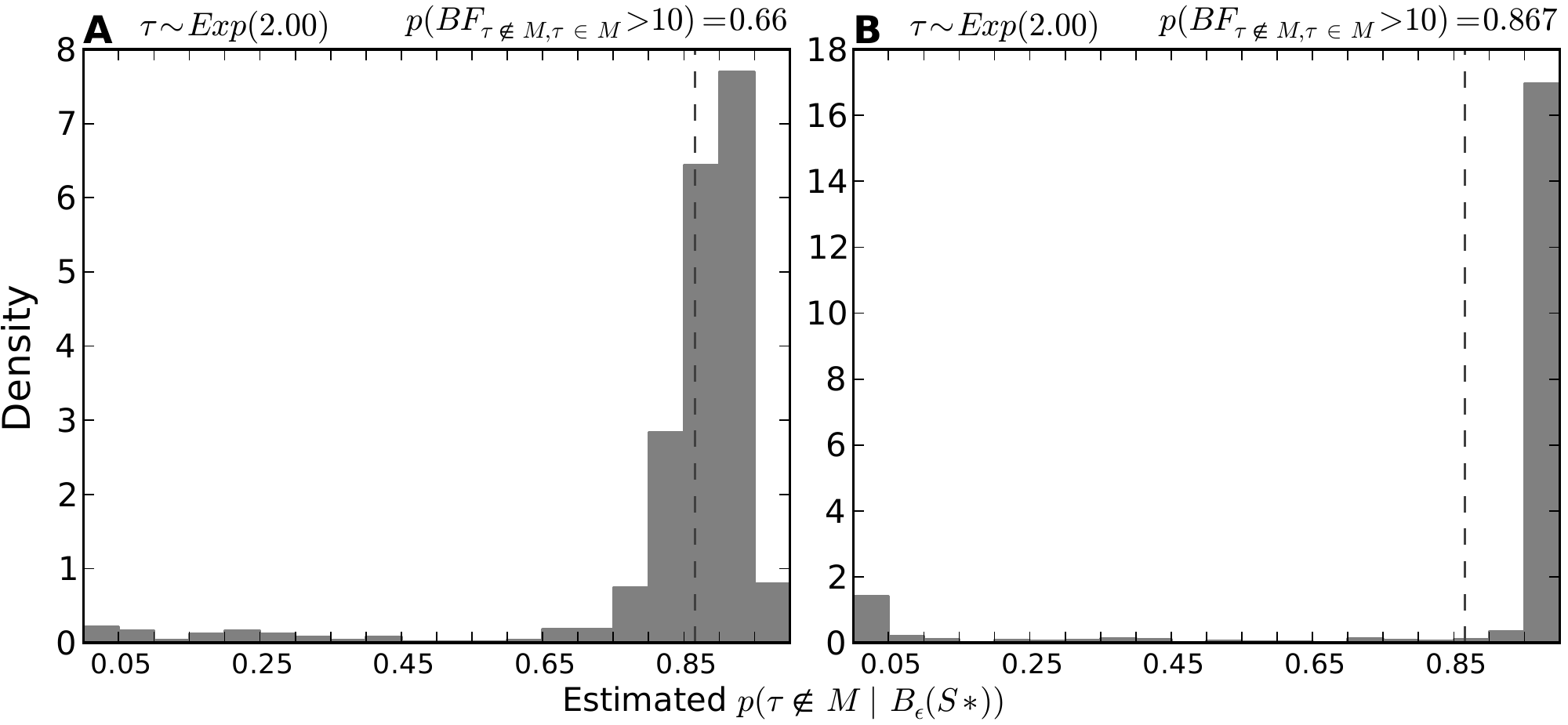}{
    Histograms of the support (estimated posterior probabilities) for excluding
    at least one true divergence time when the empirically informed
    model-averaging approach of \citet{Hickerson2013} is applied to simulated
    datasets in which divergence times of 22 pairs of populations are drawn
    from an exponential distribution, $\divt{} \sim Exp(2)$.
    The plots represent (A) unadjusted and (B) GLM-adjusted estimates from 1000
    simulation replicates analyzed using $5\e6$ samples from the prior.
    The proportion of simulation replicates in which there is strong support
    for at least one true parameter value being excluded from the model
    ($p(BF_{\divt{} \notin M, \divt{} \in M} > 10)$) is also given.
}{figExclusionSimProb}

\mFigure{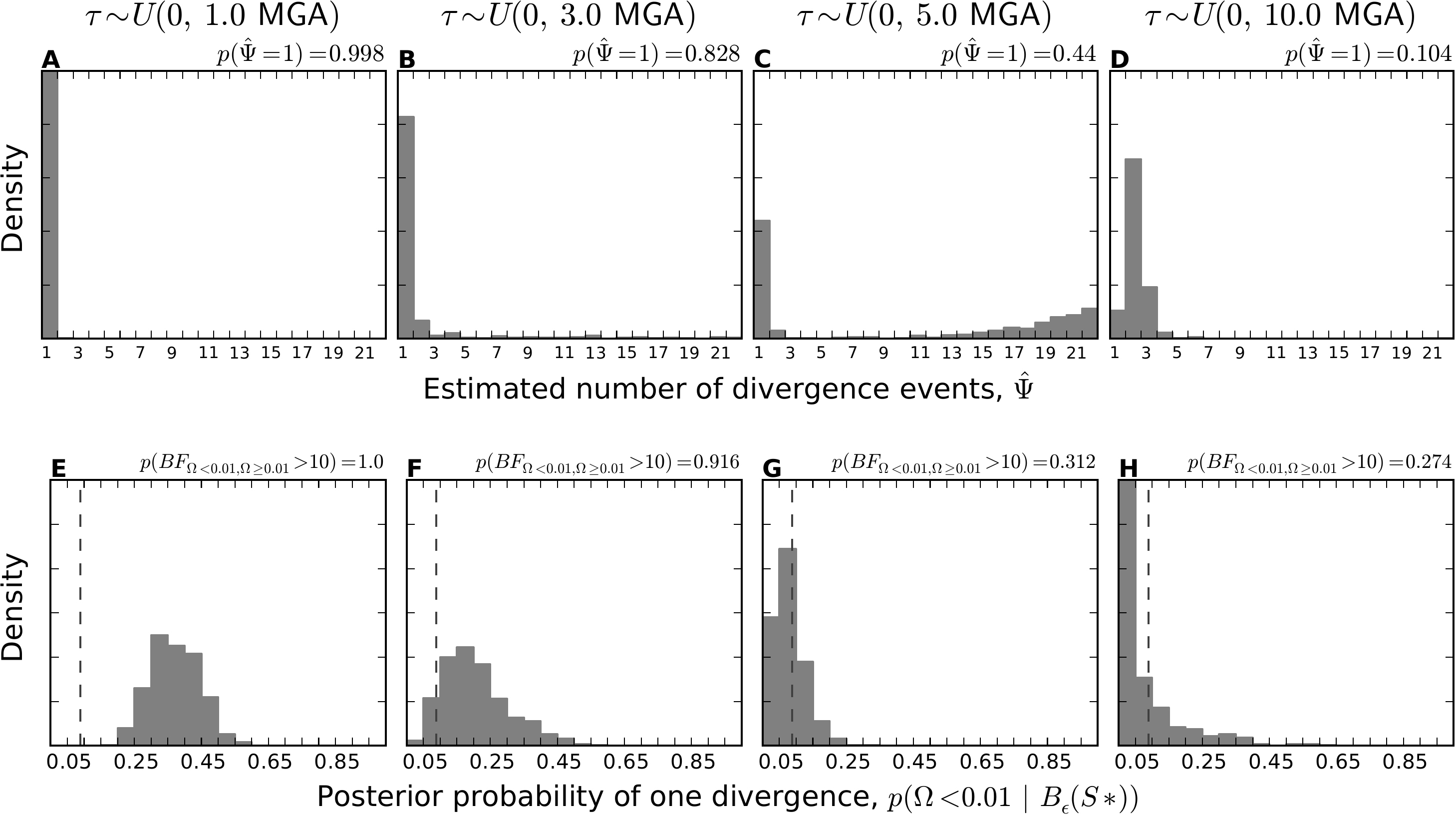}{
    The tendency of the empirically informed model-averaging approach of
    \citet{Hickerson2013} to (A--D) infer clustered divergences and (E--H)
    support the extreme model of one divergence when applied to simulated
    datasets in which the divergence times of 22 pairs of populations are
    randomly drawn from the uniform distributions $\divt{} \sim U(0,
    \divt{max})$ indicated at the top of each column of plots (divergence-time
    distributions are given in units of millions of generations ago (MGA)
    assuming a per-site rate of 1\e{-8} mutations per generation).
    Four of the six \divt{max} we simulated are provided; please see
    Figure~S\ref{figPower6} for a summary of all of the results.
}{figPower4}

\mFigure{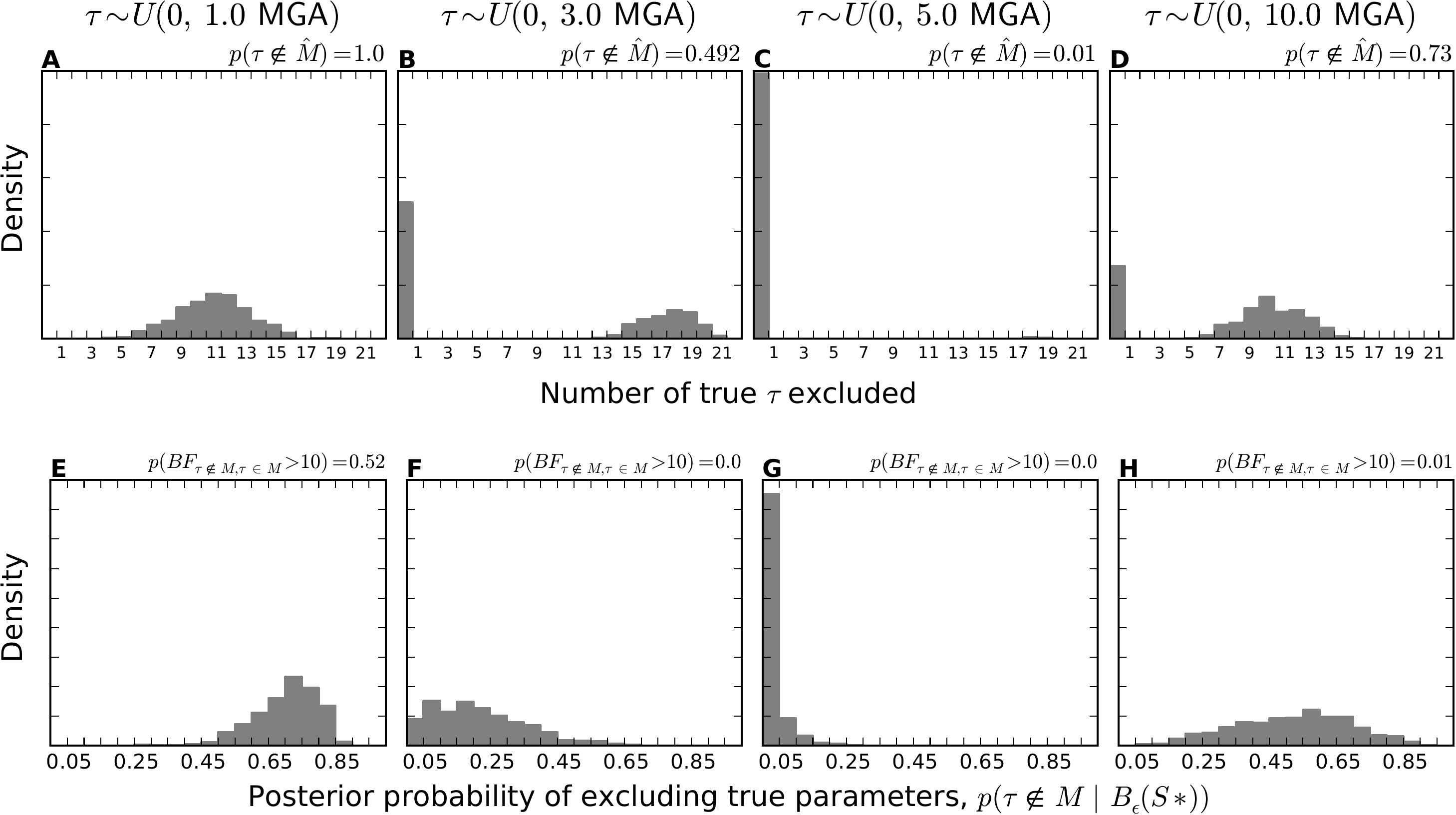}{
    Histograms of the (A--D) number of true divergence times excluded
    from the preferred model and the (E--H) posterior probability of excluding
    at least one true divergence time when the empirically informed
    model-averaging approach of \citet{Hickerson2013} is applied to simlated
    datasets in which divergence times of 22 pairs of populations are randomly
    drawn from the uniform distributions $\divt{} \sim U(0, \divt{max})$
    indicated at the top of each column of plots (divergence-time distributions
    are given in units of millions of generations ago (MGA) assuming a per-site
    rate of 1\e{-8} mutations per generation).
    Four of the six \divt{max} we simulated are provided; please see
    Figure~S\ref{figPowerExclusion6} for a summary of all of the results.
}{figPowerExclusion4}

\clearpage
\setcounter{table}{0}
\setcounter{figure}{0}
\setcounter{figure}{0}
\setcounter{table}{0}
\setcounter{page}{1}
\setcounter{section}{0}

\singlespacing

\section*{Supporting Information}
\hangindent=1cm
Oaks, J.~R., C.~W.\ Linkem, and J.~Sukumaran. \msTitle

\section{An error in Hickerson et al.'s re-analysis of the Philippines data}
\citet{Hickerson2013} re-analyzed the dataset of \citet{Oaks2012} using a
model-averaging approach, where they placed a discrete uniform prior over eight
different prior models (see Table 1 of \citet{Hickerson2013}).
However, there was an error in their methodology; their model mixes different
units of time.

Each of the eight prior models used in the re-analysis by \citet{Hickerson2013}
has one of two priors on the mean size of the descendant populations of each
taxon pair:
$\meanDescendantTheta{} \sim U(0.0001, 0.1)$ or
$\meanDescendantTheta{} \sim U(0.0005, 0.04)$.
As described in \citet{Oaks2012}, the divergence-time parameters in the model
implemented in \msb are in generations scaled relative to a constant
reference-population size, \myTheta{C}.
This reference-population size is defined in terms of the upper limit of the
uniform prior on the mean size of the descendant populations,
\meanDescendantTheta{}, such that for the prior $\meanDescendantTheta{} \sim
U(\uniformMin{\meanDescendantTheta{}},\uniformMax{\meanDescendantTheta{}})$,
the size of the constant reference population is $\myTheta{C} =
\uniformMax{\meanDescendantTheta{}}/2$.
Thus, the model used by \citet{Hickerson2013} mixes two different units of
time.
In other words, some of their prior and posterior samples are in units of
$0.05/\mutationRate$ generations, whereas others are in units of
$0.02/\mutationRate$ generations.

A fundamental assumption of the \msb model and post hoc regression adjustment
is that all possible values of the parameter of interest (divergence times) are
in the same units.
Thus, the results in sections ``Using ABC Model Comparison to Weight
Alternative Priors for the Philippine Vertebrate Data'' and ``Improved Sampling
Efficiency by Prior Weighting Supports Asynchronous and Recent Divergence for
the Philippines Vertebrate Data'' and presented in Figure 2 of
\citet{Hickerson2013} are invalid and should be disregarded.
The error is easily illustrated by re-plotting their results with the different
time units indicated (Figure~S\ref{figJointPosteriorHickerson}).

\section{Theoretical implications of empirical priors for Bayesian model
choice---A simple example}
\begin{linenomath}
The distinctions between Bayesian parameter estimation and model choice
discussed in the main text can be illustrated with a simple example.
Let us say we are interested in the fairness of a particular coin, and we
denote the unknown probability of it landing heads as \myTheta{}.
More specifically, we are interested in the probability of two models, $M_1$
and $M_2$.
In both models the outcomes of flipping the coin are assumed to be binomially
distributed, but under $M_1$ the coin is weighted toward landing heads (i.e.,
$\myTheta{} > 0.5)$), whereas under $M_2$, the coin is weighted toward landing
tails (i.e., $\myTheta{} < 0.5$).
We already have data from flipping a different coin 20 times that landed both
heads and tails 10 times each, and so we decide to use these data in specifying
a beta prior on fairness of the new coin of $beta(a=10, b=10)$
(Figure~S\ref{figCoinFlip}).
We collect data by flipping the coin of interest $N=10$ times, $y=3$ of which
land heads.
Given the beta distribution is a conjugate prior for a binomial likelihood, the
posterior distribution has the nice analytical form $\theta \given y,N \sim
beta(a + y, b + N - y)$, which for the new dataset is simply $beta(13, 17)$
(Figure~S\ref{figCoinFlip}).
The maximum a posteriori (MAP) estimate of the probability of heads is 0.429,
and following Equation~\ref{eq:marginallikelhiood} in the main text the
marginal likelihoods of our models of interest are
\begin{equation}
    p(y=3, N=10 \given M_1) = \int_{0.5}^{1} p(y=3, N=10 \given
    \myTheta{}, M_1)p(\myTheta{} \given M_1) d\myTheta{} \approx 0.029,
\end{equation}
and
\begin{equation}
    p(y=3, N=10 \given M_2) = \int_{0}^{0.5} p(y=3, N=10 \given
    \myTheta{}, M_2)p(\myTheta{} \given M_2) d\myTheta{} \approx 0.097.
\end{equation}
Given the models have equal probability under our prior, we can calculate the
posterior probability of Model 1 as
\begin{equation}
    p(M_1 \given y=3, N=10) = \frac{p(y=3, N=10 \given M_1)}{p(y=3, N=10 \given
    M_1) + p(y=3, N=10 \given M_2)} \approx 0.23.
\end{equation}
This is the correct posterior probability of Model 1 given our prior
and data.
\end{linenomath}

To give the data more weight relative to the prior, we could use it twice, and
calculate an empirical Bayes estimate using a prior of $beta(13,17)$.  This
results in a ``posterior'' distribution of $beta(16, 24)$
(Figure~S\ref{figCoinFlip}), with a MAP estimate of 0.395, and $p(M_1 \given
y=3, N=10) = 0.10$.
The estimated posterior distribution of the parameter, and resulting MAP
estimate, is similar whether or not an empirically informed prior is used.
However, the posterior probability of Model 1 is very sensitive to the
empirical prior, decreasing by 56\%.
By using the empirically informed prior, we ignored prior uncertainty, leading
to an underestimate of our posterior uncertainty (Figure~S\ref{figCoinFlip}).
While this did not greatly affect our estimate of \myTheta{}, it misled us
to be overconfident in Model 2.

\section{Validation analyses}
Following \citet{Oaks2012}, we characterize the model-choice behavior of the
model-averaging approach of \citet{Hickerson2013} under the ideal conditions
where the prior is correct (i.e., the data are generated from parameters drawn
from the same prior distributions used in the analysis).
We used the same prior models as above ($M_1$--$M_5$;
Table~\ref{tabModelChoiceEmpirical}), and simulated 50,000 datasets under this
prior (10,000 from each model).
We used a simulated data structure of eight population pairs, with a single
1000 base-pair locus sampled from 10 individuals from each population.
We then analyzed each of these replicate datasets using the same prior with 2.5
million samples (500,000 from each of the five prior models), retaining 1000
posterior samples.
Our results are very similar to \citet{Oaks2012}, but we note that they
are not directly comparable as our simulations contained eight population
pairs rather than 10 (Figure~\ref{figValidationMCBehavior}).
We find that the approach of \citet{Hickerson2013} estimates the posterior
probability of divergence models reasonably well when all assumptions of the
method are met (i.e., the prior is correct) and the unadjusted posterior
estimates are used.
Similar to \cite{Oaks2012}, we find that the regression-adjusted estimates of
the model probabilities are biased.

\section{A difficult inference problem}
In the main text, we discuss how the prior assumption of uniformly distributed
divergence times in \msb leads to posteriors that are difficult to interpret.
However, it is also important to consider the difficult inference problem with
which \msb is faced.
When applying \msb to the dataset of \citet{Oaks2012} with 22 taxon
pairs, there are 581--602 free parameters that model highly stochastic
coalescent and mutational processes.
Under this rich stochastic model, the method is estimating the
probability of 1002 divergence models \citep[i.e., the number of integer
partitions of $Y=22$;][]{Oaks2012}.
Furthermore, all the information in the sequence alignment of each taxon pair
is distilled into four summary statistics.
This gives us a total of 88 summary statistics (four from each of the 22 taxon
pairs) that contain minimal information about many of the $\approx 600$
parameters in the model.
More summary statistics can be used in \msb, but most are highly correlated
with the four default statistics, and thus contribute little additional
information about the parameters from the sequence data.
The large number of parameters and divergence models relative to the amount of
information in the data is undoubtedly another reason the method lacks
robustness to prior conditions.

\section{Additional clarifications from \citet{Hickerson2013}}

\subsection{Saturation of summary statistics}
\citet{Hickerson2013} claim the priors used by \citet{Oaks2012} ``cause much of
the explored parameter space to be beyond the threshold of saturation in most
mtDNA genes.'' To explore this possibility, we simulated datasets under prior
settings that match two of the three priors used by \citet{Oaks2012}:
$\meanDescendantTheta{} \sim U(0.0005, 0.04)$ and $\ancestralTheta{} \sim
U(0.0005, 0.02)$.
Under this prior, we randomly sample divergence-time parameters from a uniform
distribution of $U(0, 20)$ coalescent units, simulate datasets, and plot the
\divt{} values against the summary statistics calculated from the resulting
datasets (Figure~\ref{figSaturationPlot}).
Clearly, the priors used by \citet{Oaks2012} with upper limits on \divt{} of five
and 10 coalescent units suffered little to no effect from saturation.
Even at divergence times of 20 coalescent units, there is still signal in the
summary statistics used by \msb (Figure~\ref{figSaturationPlot}).
Thus, the assertion of \citet{Hickerson2013} that the priors used by
\citet{Oaks2012} sample parameter space in which the mtDNA alignments are
saturated by substitutions is incorrect and, as a result, does not explain the
bias they found.

\subsection{Graphical prior comparisons}
\citet{Hickerson2013} advocate the use graphical checks of prior models.
This prior-predictive approach entails generating a small number (1000) of
random samples from the prior and plotting the resulting summary statistics in
comparison to the observed statistics to see if they coincide (see Figure 1 of
\citet{Hickerson2013}).
Given the richness of the \msb model ($\approx 600$ parameters for the Philippine
dataset analyzed by \citet{Hickerson2013}), we do not expect that 1000
\emph{random} draws from the vast prior parameter space will yield data and
summary statistics consistent with the observed data.
In fact, when such random draws are tightly clustered around the observed
statistics, this can be an indication that the prior is over-fit, as we show in
the main text (Table~\ref{tabModelChoiceEmpirical} and Figure~S\ref{figPCA}).
Thus, using such plots to select priors should be avoided, and the use of
posterior-predictive analyses would be much more informative about the overall
fit of models.

\subsection{Differing utilities of \numt{} and \vmratio{} in \texttt{msBayes}}
The primary component of the \msb model is the vector of divergence
times for each of the taxon pairs,
$\divtvector = \{\divt{1}, \ldots, \divt{Y}\}$
\citep{Oaks2012}.
\citet{Hickerson2013} argue that the dispersion index of this vector,
\vmratio{}, is a better model-choice estimator than the number of 
divergence-time parameters within the vector,
\numt{}.
They present a plot of \numt{} against \vmratio{} (Fig.~S1 of
\citet{Hickerson2013}), which is essentially a plot of sample size versus
variance.
This plot shows that \vmratio{} has very little information
about the number of divergences among taxa.
Nonetheless, \citet{Hickerson2013} conclude \vmratio{} is more informative and
biogeographically relevant than \numt{}.
However, the number of divergence-time parameters within the vector and their
values contains all of the information about the temporal distribution of
divergences, and is much more informative than the variance (i.e., the
dispersion index is not a sufficient statistic for \divtvector).
\citet{Hickerson2013} also argue that \msb can estimate \vmratio{} much better
than \numt{}.
However, \citet{Oaks2012} demonstrate that even when all assumptions of the
model are met, \vmratio{} is a poor model-choice estimator (see plots B, D \& F
of Figure 4 in \citet{Oaks2012}), whereas \numt{} performs better.

Importantly, \vmratio{} is limited to estimating the probability of only a
single model (the one-divergence model), and thus its utility for model-choice
is very limited.
I.e., it can only be informative about the probability of whether there is one
divergence shared among the taxa ($\vmratio{} = 0.0$) or there is greater than
one divergence ($\vmratio{} > 0.0$).
As a result, not only is its model-choice utility limited, but it is also
very difficult to estimate.
\vmratio{} can range from zero to infinity, and the point density that it is
at its lower limit of zero will always be zero.
Thus, an arbitrary threshold (0.01 is used throughout the \msb literature) must
be chosen to make the probability of ``simultaneous'' divergence estimable.
Even with this arbitrary threshold, it is still not surprising to see that it
is numerically difficult to obtain reliable estimates of the probability that
\vmratio{} is ``near'' its lower limit of zero.
It is easier, less subjective, and more interpretable to estimate the
probability of the model with one divergence-time parameter (i.e., $\numt{} =
1$).
Thus, it is not surprising that \citet{Oaks2012} find that \numt{} is a better
estimator of model probability than \vmratio{}.

\newpage
\singlespacing

\siFigure{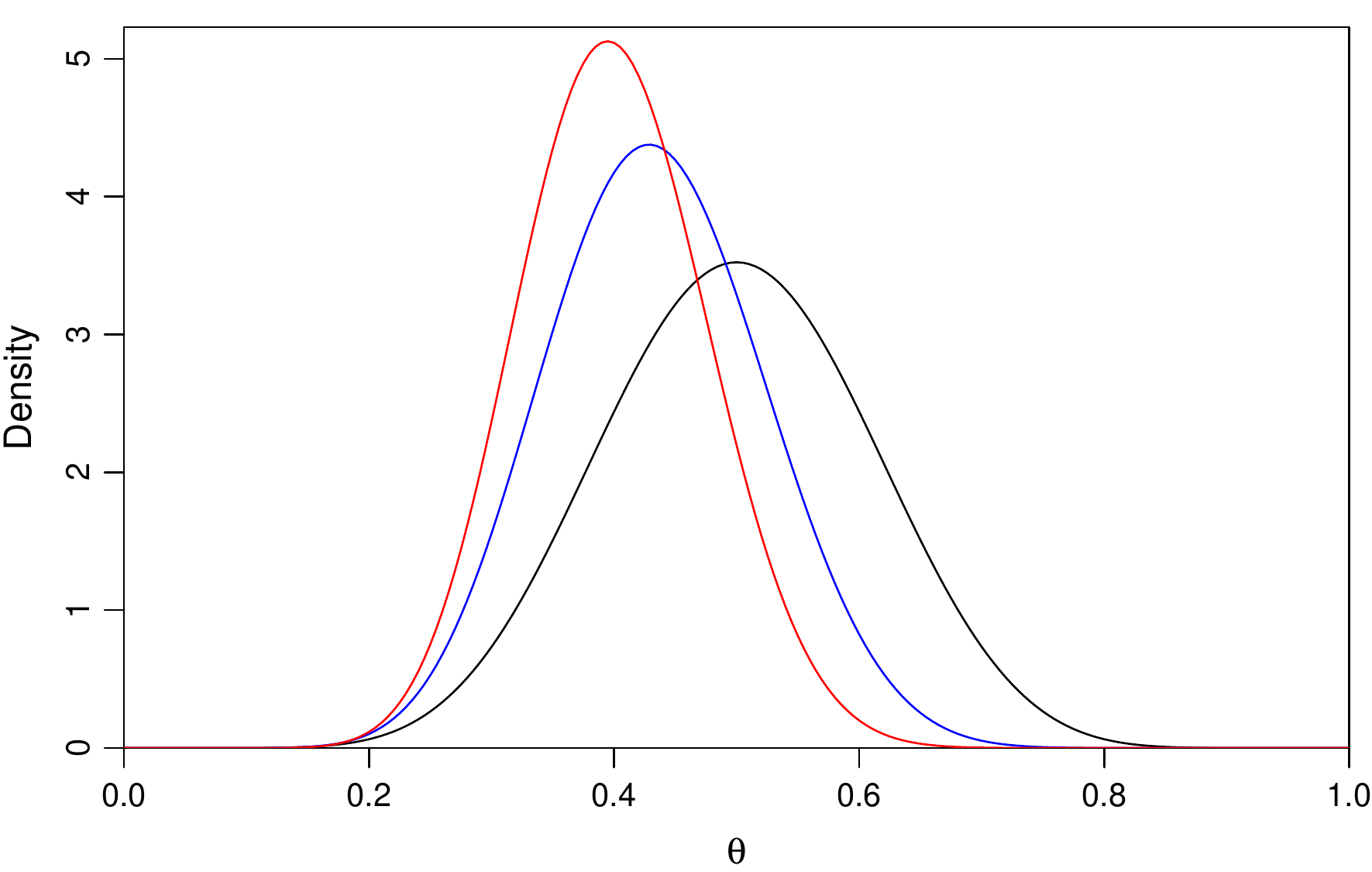}{
    A plot of three beta probability density functions that represent a prior
    (black; $beta(10, 10)$), posterior (blue; $beta(13, 17)$), and empirical
    Bayes density (red; $beta(16, 24)$) for a dataset of 10 coin flips, three
    of which are successes.
}{figCoinFlip}

\siFigure{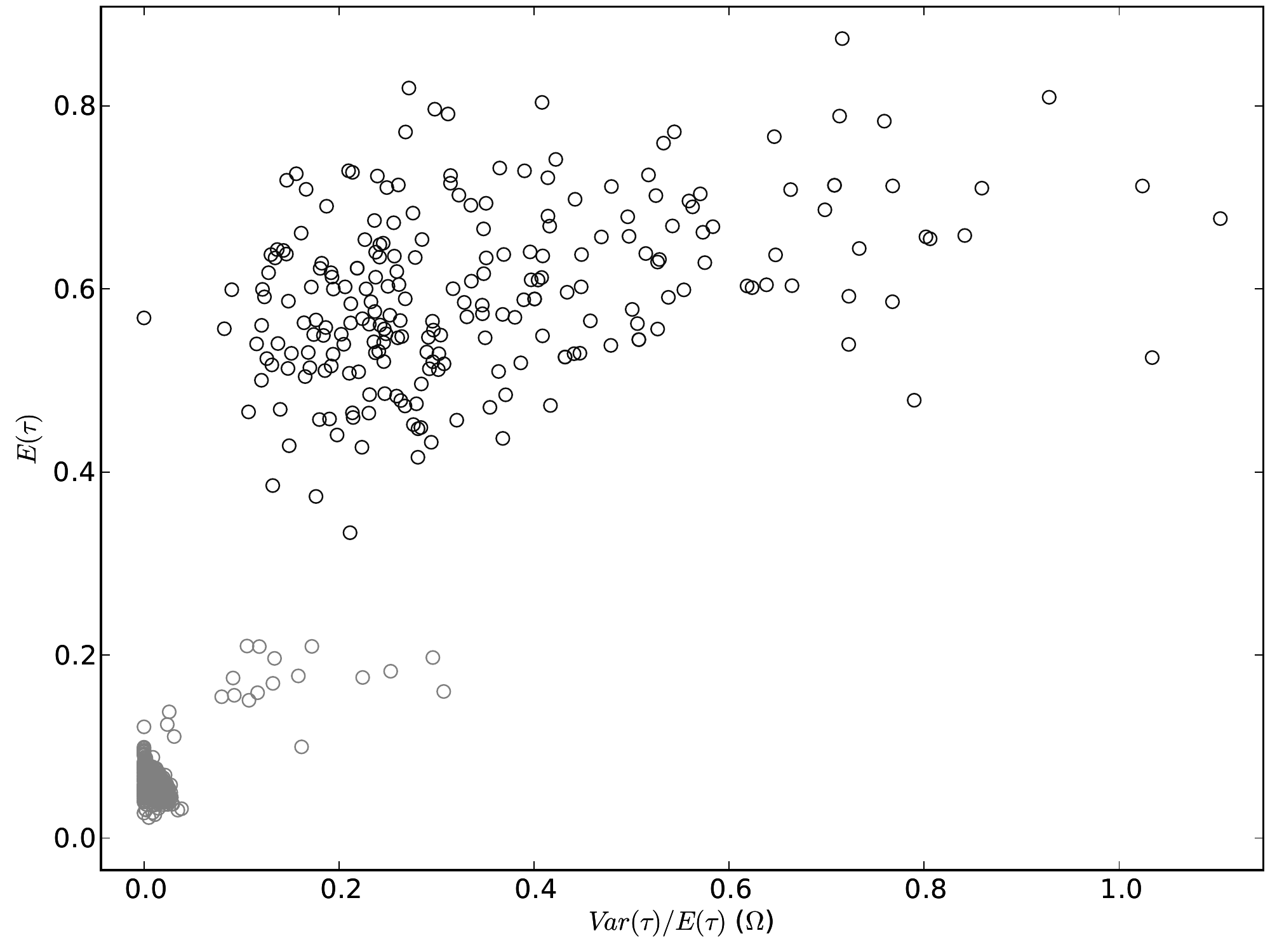}{
    The joint posterior of the mean (\meant{}{}) and dispersion index ($\vmratio{} = 
    \vart{}{}/\meant{}{}$) of divergence times for 22 vertebrate taxon pairs as
    estimated by \citet{Hickerson2013} (see Figure 2B of \citet{Hickerson2013}).
    The posterior samples are color-coded to indicate the erroneous mixture of
    timescales in the analysis of \citet{Hickerson2013};
    grey = $0.05/\mutationRate$ generations and
    black = $0.02/\mutationRate$ generations.
}{figJointPosteriorHickerson}

\siFigure{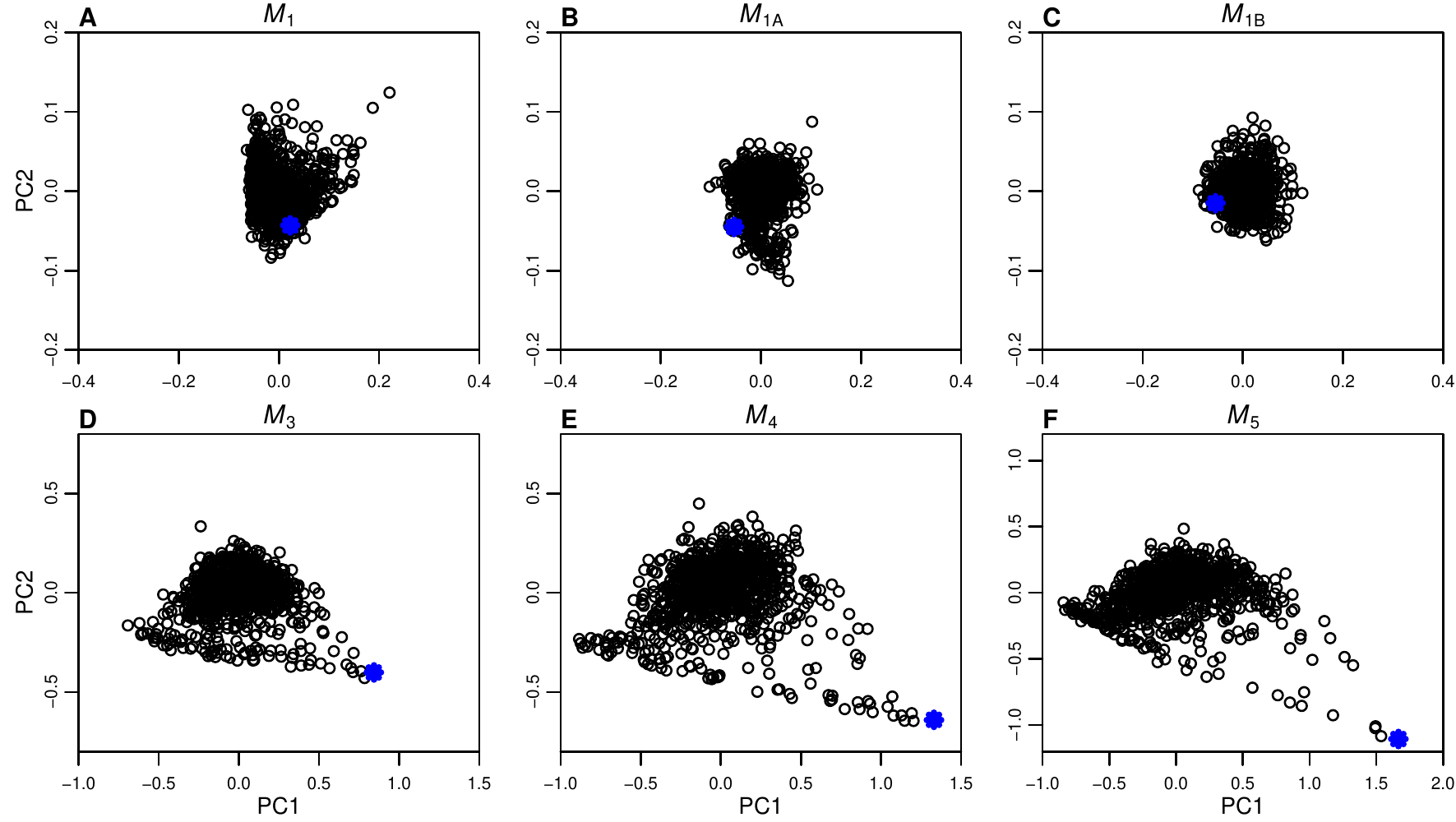}{
    The prior predictive graphical checks recommended by \citet{Hickerson2013}
    for six prior models:
    (A) $M_1$ ($\divt{} \sim U(0, 0.1)$),
    (B) $M_{1A}$ ($\divt{} \sim U(0, 0.01)$),
    (C) $M_{1B}$ ($\divt{} \sim U(0, 0.001)$), 
    (D) $M_3$ ($\divt{} \sim U(0, 5)$),
    (E) $M_4$ ($\divt{} \sim U(0, 10)$), and
    (F) $M_5$ ($\divt{} \sim U(0, 20)$).
    The three models that likely exclude true values of some divergence times
    of the 22 pairs of Philippine taxa (A--C) appear to have a ``better fit''
    than the valid priors that likely cover the true divergence times
    (D--F).
    The plots project the summary statistics from 1000 random samples from each
    model onto the first two orthogonal axes of a principle component analysis,
    with the blue dot representing the observed summary statistics from the 22
    population pairs of Philippine vertebrates.
}{figPCA}

\siSidewaysFigure{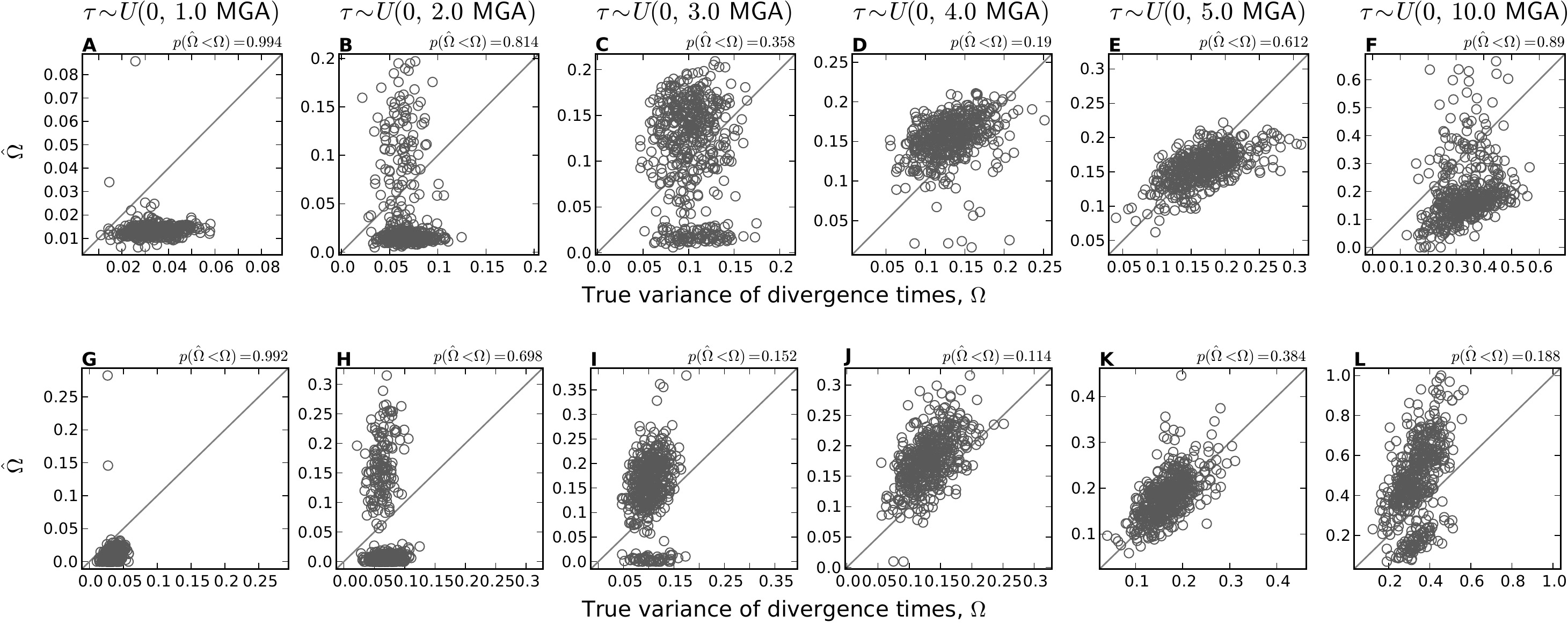}{
    The accuracy of (A--F) unadjusted and (G--L) GLM-adjusted estimates of
    the dispersion index of divergence times (\vmratio{}) when the empirically
    informed model-averaging approach of \citet{Hickerson2013} is applied to
    simlated datasets in which divergence times of 22 pairs of populations are
    randomly drawn from the uniform distributions $\divt{} \sim U(0,
    \divt{max})$ indicated at the top of each column of plots (divergence-time
    distributions are given in units of millions of generations ago (MGA)
    assuming a per-site rate of 1\e{-8} mutations per generation).
}{figPowerAccuracy}

\siSidewaysFigure{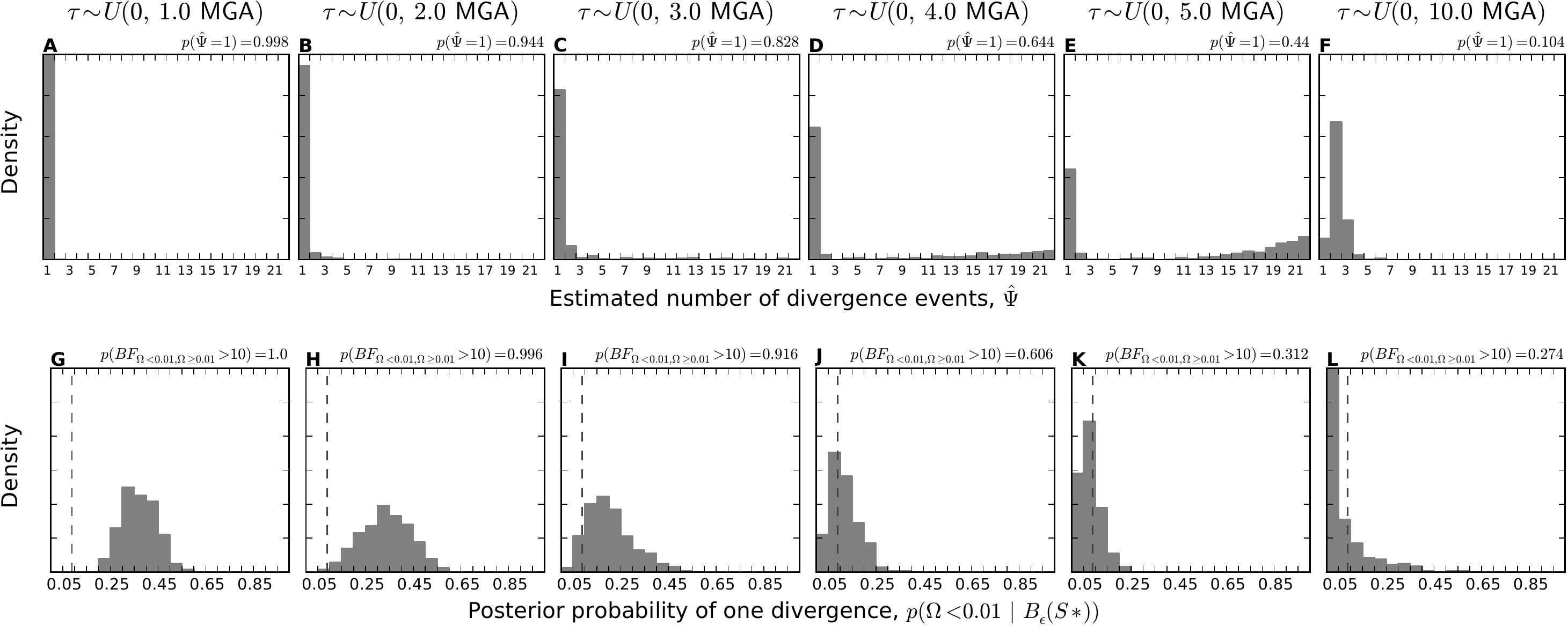}{
    The tendency of the empirically informed model-averaging approach of
    \citet{Hickerson2013} to (A--F) infer clustered divergences and (G--L)
    support the extreme model of one divergence when applied to simulated
    datasets in which the divergence times of 22 pairs of populations are
    randomly drawn from the uniform distributions $\divt{} \sim U(0,
    \divt{max})$ indicated at the top of each column of plots (divergence-time
    distributions are given in units of millions of generations ago (MGA)
    assuming a per-site rate of 1\e{-8} mutations per generation).
}{figPower6}

\siSidewaysFigure{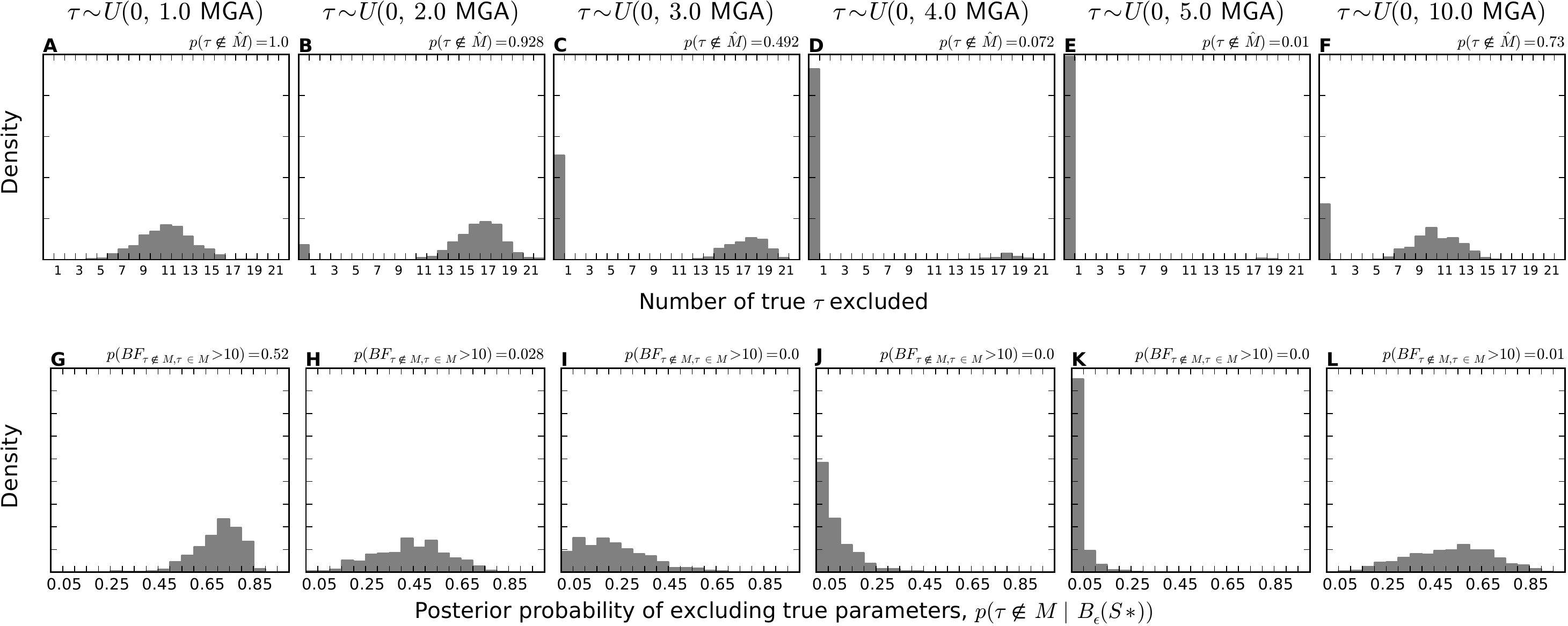}{
    Histograms of the (A--F) number of true divergence times excluded
    from the preferred model and the (G--L) posterior probability of excluding
    at least one true divergence time when the empirically informed
    model-averaging approach of \citet{Hickerson2013} is applied to simlated
    datasets in which divergence times of 22 pairs of populations are randomly
    drawn from the uniform distributions $\divt{} \sim U(0, \divt{max})$
    indicated at the top of each column of plots (divergence-time distributions
    are given in units of millions of generations ago (MGA) assuming a per-site
    rate of 1\e{-8} mutations per generation).
}{figPowerExclusion6}

\widthFigure{0.9}{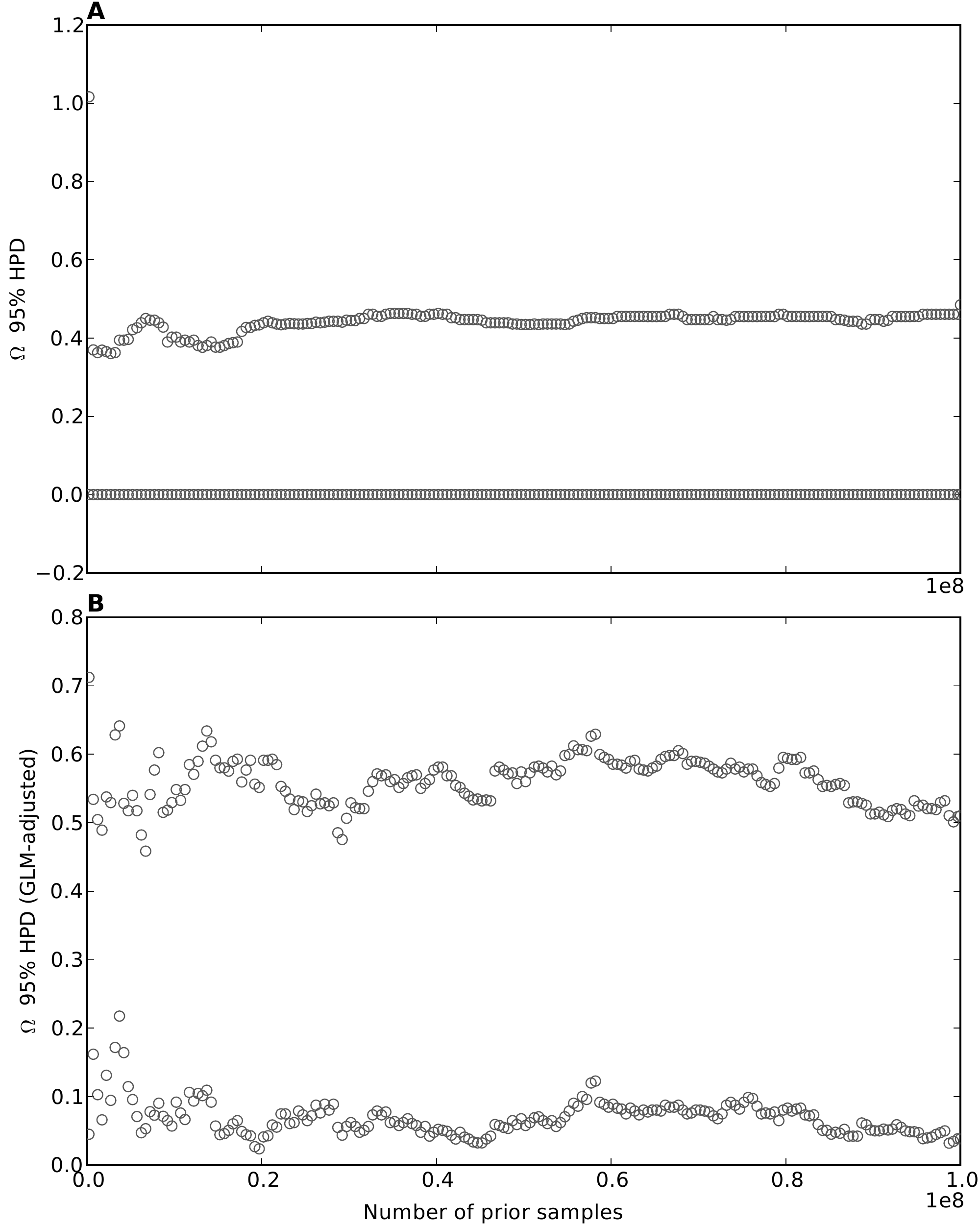}{name=Figure S, labelformat=noSpace, listformat=sFigList}{
    Traces of the estimated lower and upper limits of the 95\% highest posterior density (HPD)
    interval of \vmratio{} (the dispersion index of divergence times) as 100 million prior
    samples are accumulated. Each pair of points is based on 1000 posterior samples retained
    from the prior. Both (A) unadjusted and (B) GLM-regression-adjusted estimates are shown.
    The data analyzed were the 22 pairs of Philippine taxa from \citet{Oaks2012}.
    Prior settings were $\divt{} \sim U(0,10)$, $\meanDescendantTheta{} \sim U(0.0005, 0.04)$,
    and $\ancestralTheta{} \sim U(0.0005, 0.02)$.
}{figSamplingError}
\clearpage

\siFigure{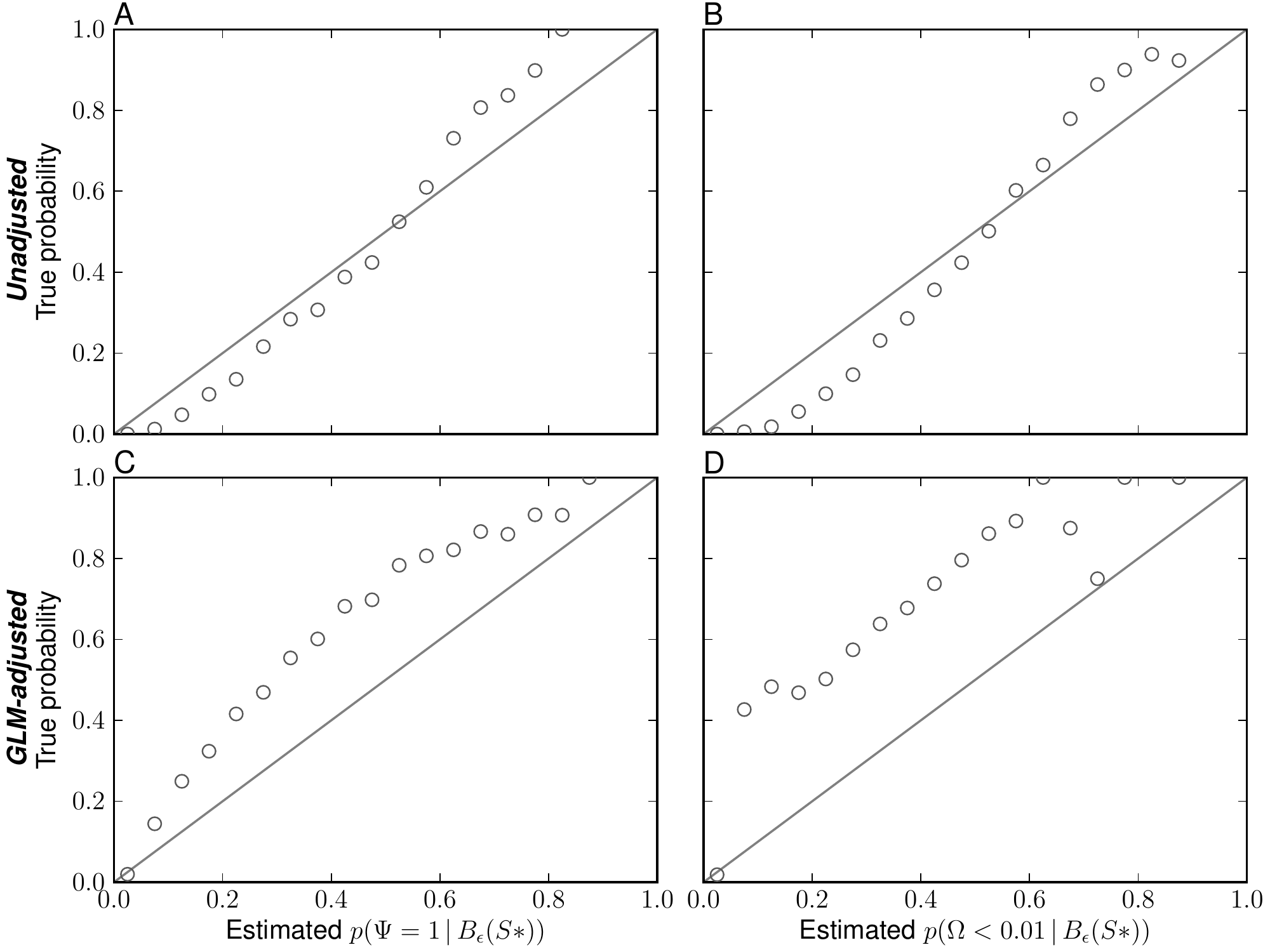}{
    An assessment of the approximate Bayesian model-averageing approach of
    \citet{Hickerson2013} under the ideal conditions when the prior model
    is correct (i.e., the datasets are simulated from parameters drawn from the
    same prior distributions used in the analysis).
    The plots show the relationship between the estimated posterior and true
    probability of (A \& C) $\numt{}=1$ and (B \& D) $\vmratio{} < 0.01$, based
    on 50,000 simulations.
    The results summarize the (A \& B) unadjusted and (C \& D) GLM-adjusted
    posterior estimate from each simulation replicate.
    The prior settings for all replicates included five prior models with
    $\meanDescendantTheta{} \sim U(0.0001, 0.1)$ and $\ancestralTheta{} \sim
    U(0.0001, 0.05)$ for all five models, and
    $M_1: \divt{} \sim U(0, 0.1)$,
    $M_2: \divt{} \sim U(0, 1)$,
    $M_3: \divt{} \sim U(0, 5)$,
    $M_4: \divt{} \sim U(0, 10)$, and
    $M_5: \divt{} \sim U(0, 20)$.
    The number of samples from the prior was $2.5\e6$.
    The simulated data structure was 8 population pairs, with a single 1000
    bp locus sampled from 10 individuals from each population.  The 50,000
    estimates of the posterior probability of one divergence event were
    assigned to 20 bins of width 0.05.
    The estimated posterior probability of each bin is plotted against the
    proportion of replicates in that bin with a true value consistent with
    one divergence event (i.e., $\numt{}=1$ or $\vmratio{} < 0.01)$.
}{figValidationMCBehavior}

\siFigure{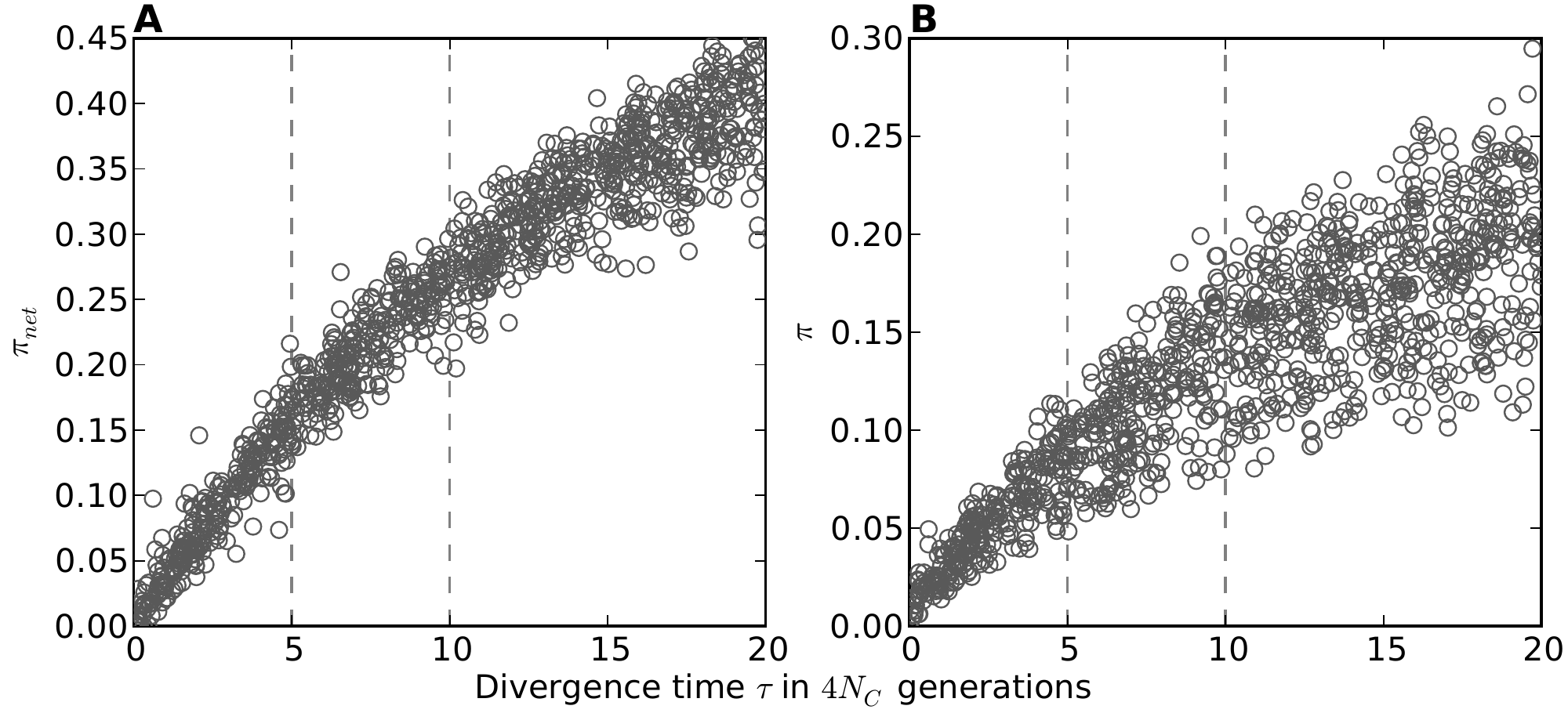}{
    The summary statistics $\pi$ \citep{Tajima1983} and $\pi_{net}$
    \citep{Takahata1985} as a function of divergence time between populations.
    Each plot represents 1100 pairs of parameter draws and summary statistics
    calculated from the simulated data.
    Prior settings for the simulations were $\divt{} \sim U(0, 20)$,
    $\meanDescendantTheta{} \sim U(0.0005, 0.04)$, and $\ancestralTheta{} \sim
    U(0.0005, 0.02)$.
}{figSaturationPlot}

\siFigure{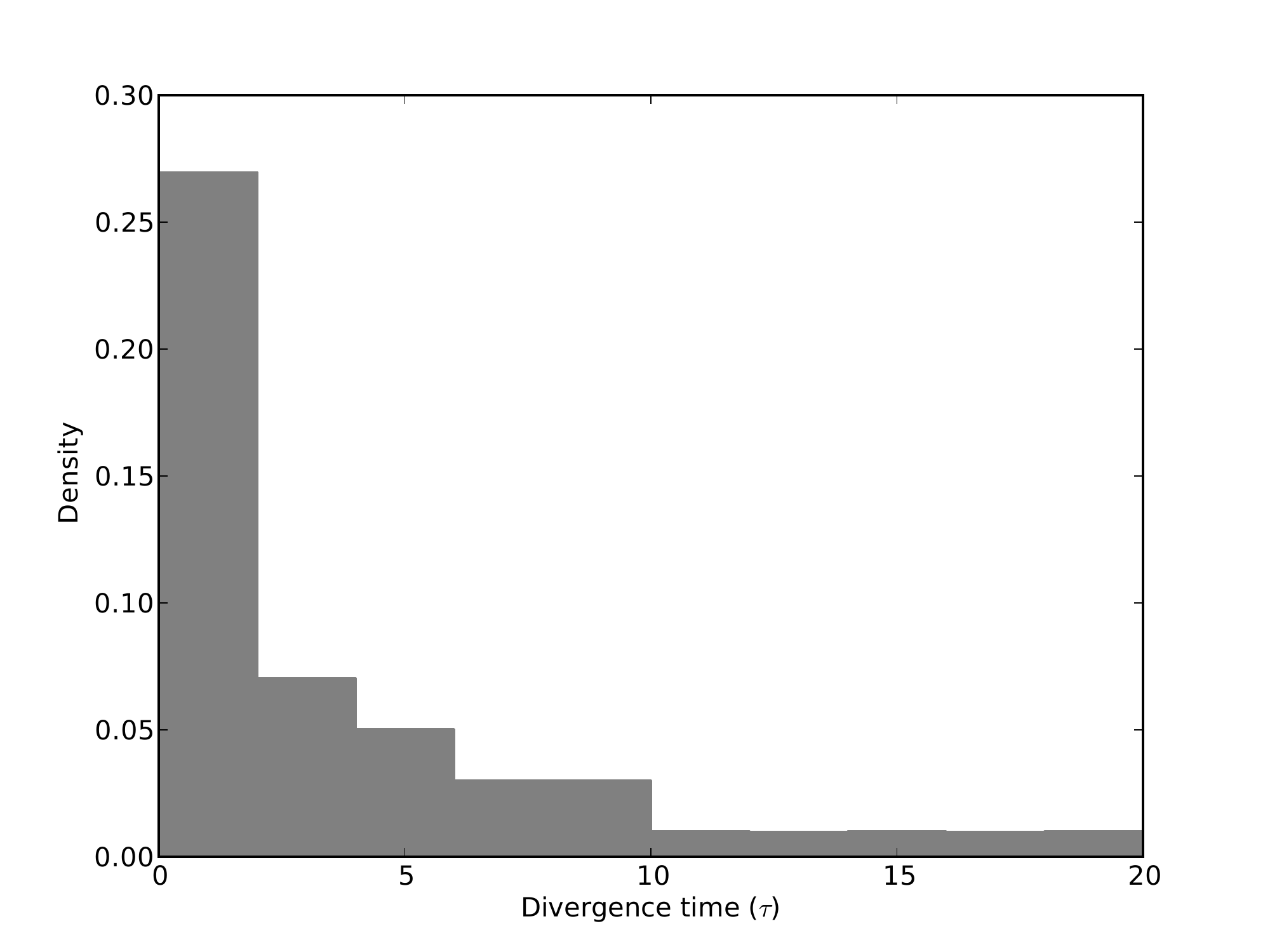}{
    The prior distribution on divergence times imposed by the model-averaging prior
    comprised of five models with different uniform priors on \divt{}:
    $M_1$ ($\divt{} \sim U(0, 0.1)$), $M_2$ ($\divt{} \sim U(0, 1)$), $M_3$
    ($\divt{} \sim U(0, 5)$), $M_4$ ($\divt{} \sim U(0, 10)$), $M_5$ ($\divt{}
    \sim U(0, 20)$).
}{figMCTauPrior}

\end{document}